\documentclass[twocolumn,amssymb,superscriptaddress,showpacs]{iopart}

\usepackage{iopams}

\expandafter\let\csname equation*\endcsname\relax

\expandafter\let\csname endequation*\endcsname\relax

\usepackage{amsmath}
\usepackage[latin9]{inputenc}
\pdfoutput=1
\usepackage[english]{babel}
\usepackage[T1]{fontenc}
\usepackage{amsmath}
\usepackage{hyperref}
\usepackage{dsfont}
\usepackage{graphicx}
\usepackage{appendix}
\usepackage{amssymb}

\newcommand{\bra}[1]{\langle #1|}
\newcommand{\ket}[1]{|#1\rangle}
\newcommand{\ketbra}[1]{| #1\rangle \langle #1|}

\newcommand{\be}{\begin{equation}}
\newcommand{\ee}{\end{equation}}
\newcommand{\eea}{\end{eqnarray}}
\newcommand{\bea}{\begin{eqnarray}} 

\newcommand{\va}[1]{\ensuremath{(\Delta#1)^2}}

\newcommand{\ex}[1]{\ensuremath{\langle{#1}\rangle}}

\newcommand{\qed}{\ensuremath{\hfill \blacksquare}}

\newcommand{\kommentar}[1]{}

\newcommand{\forget}[1]{}

\newcommand{\FQ}{\mathcal F_{Q}}

\newcommand{\EQ}[1]{equation~\eqref{#1}}
\newcommand{\EQS}[1]{equations~\eqref{#1}}
\newcommand{\EQL}[1]{Equation~\eqref{#1}}

\newcommand{\FIG}[1]{figure~\ref{#1}}
\newcommand{\REF}[1]{\cite{#1}}

\newcommand{\RES}[1]{Result~\ref{#1}}

\usepackage{orcidlink} 
\usepackage{cite}
\newcommand{\app}[1]{\ref{#1}}
\newcommand{\SUPP}[1]{\ref{#1}}

\usepackage{ntheorem}
\theoremheaderfont{\hspace*{\parindent}\bf}
\theoremseparator{.}
\theorempreskip{0cm}
\theorempostskip{0cm}
\newtheorem{result}{Result}
\let\oldresult\result
\renewcommand{\result}{\oldresult\normalfont}

\newcounter{mysection}
\makeatletter
\@addtoreset{section}{mysection}
\makeatother

\newcounter{myequation}
\makeatletter
\@addtoreset{equation}{myequation}
\makeatother

\newcounter{myfigure}
\makeatletter
\@addtoreset{figure}{myfigure}
\makeatother

\newcounter{myresult}
\makeatletter
\@addtoreset{result}{myresult}
\makeatother

\begin{document}

\title{Activation of metrologically useful genuine multipartite entanglement}

\bigskip
\author{R\'obert Tr\'enyi$^{1,2,3,4}$\orcidlink{0000-0002-2839-0472}, \'Arp\'ad Luk\'acs$^{1,5,4}$\orcidlink{0000-0002-5737-1393},\\ Pawe{\l}  Horodecki$^{6,7}$\orcidlink{0000-0002-3233-1336}, Ryszard Horodecki$^{6}$\orcidlink{0000-0003-2935-290X},\\ Tam\'as V\'ertesi$^{8}$\orcidlink{0000-0003-4437-9414} and G\'eza T\'oth$^{1,2,3,9,4,*}$\orcidlink{0000-0002-9602-751X}}

\address{$^1$ Department of Theoretical Physics,  University of the Basque Country UPV/EHU, P.O. Box 644, E-48080 Bilbao, Spain}

\address{$^2$ EHU Quantum Center, University of the Basque Country UPV/EHU, \\Barrio Sarriena s/n, E-48940 Leioa, Biscay, Spain}

\address{$^3$ Donostia International Physics Center (DIPC), \\P.O. Box 1072, E-20080 San Sebasti\'an, Spain}

\address{$^4$ HUN-REN Wigner Research Centre for Physics,\\ P.O. Box 49, H-1525 Budapest, Hungary}

\address{$^5$ Department of Mathematical Sciences, Durham University,  \\Stockton Road, DH1 3LE Durham, United Kingdom}

\address{$^6$ International Centre for Theory of Quantum Technologies, University of Gda\'nsk, Wita Stwosza 63, 80-308 Gda\'nsk, Poland}

\address{$^7$ Faculty of Applied Physics and Mathematics, National Quantum Information Centre, Gda\'nsk University of Technology, Gabriela Narutowicza 11/12, \\80-233 Gda\'nsk, Poland}

\address{$^8$ MTA Atomki Lend\"ulet Quantum Correlations Research Group, \\ HUN-REN Institute for Nuclear Research, Hungarian Academy of Sciences, \\P.O. Box 51, H-4001 Debrecen, Hungary}

\address{$^9$ IKERBASQUE, Basque Foundation for Science,
E-48013 Bilbao, Spain}
\address{$^*$ Author to whom any correspondence should be addressed.}
\ead{toth@alumni.nd.edu}

\begin{abstract}
We consider quantum metrology with several copies of bipartite and multipartite quantum states. We characterize the metrological usefulness by determining how much the state outperforms separable states. We identify a large class of entangled states that become maximally useful for metrology in the limit of large number of copies, even if the state is weakly entangled and not even more useful than separable states. This way we activate metrologically useful genuine multipartite entanglement. Remarkably, not only that the maximally achievable metrological usefulness is attained exponentially fast in the number of copies, but it can be achieved by the measurement of few simple correlation observables. We also make general statements about the usefulness of a single copy of pure entangled states.  We surprisingly find that the multiqubit states presented in Hyllus {\it et al.} [\href{https://doi.org/10.1103/PhysRevA.82.012337}{Phys. Rev. A {\bf 82}, 012337 (2010)}], which are not useful,  become useful if we embed the qubits locally in qutrits.  We discuss the relation of our scheme to error correction, and its possible use for quantum metrology in a noisy environment.

\vspace{1em}\noindent
DOI: \href{https://doi.org/10.1088/1367-2630/ad1e93}{10.1088/1367-2630/ad1e93}
\end{abstract}

\date{\today}

\maketitle
\section{Introduction}
Quantum entanglement plays a central role in quantum physics, as well as in quantum information processing applications \cite{Horodecki2009Quantum,Guhne2009Entanglement,Friis2019}. There have been numerous experiments creating entanglement with photons, trapped cold ions, cold atoms and superconducting circuits \cite{Julsgaard2001Experimental,Esteve2008Squeezing,Behbood2014Generation,Bohnet2016Quantum,Kong2020Measurement-induced,Riedel2010Atom-chip-based,Gross2010Nonlinear,Lucke2014Detecting,McConnell2015Entanglement,Hosten2016Measurement,Zou2018Beating,Xin2023Long-Lived,Sackett2000Experimental,Leibfried2005Creation,Kiesel2005Experimental,Walther2005ExperimentalOneWay,Haffner2005Scalable,Kiesel2007Experimental,Wieczorek2009Experimental,Prevedel2009Experimental,Song201710-Qubit,Monz201114-Qubit,Huang2011Multi-partite,Wang201818-Qubit,Friis2018Observation,Gong2019Genuine,Song2019Generation,Compact2021Pogorelov,Thomas2022Efficient,Cao2023Generation}. In the multiparticle scenario, a state can be not only separable or entangled, but it can possess various levels of $k$-particle entanglement \cite{Dur2000Three,Acin2001Classification,Sorensen2001Entanglement,Guhne2005Multipartite,Vitagliano2017Entanglement,Vitagliano2018Entanglement,Riedel2010Atom-chip-based,Gross2010Nonlinear,Lucke2014Detecting,McConnell2015Entanglement,Hosten2016Measurement,Zou2018Beating,Xin2023Long-Lived}.  The highest form of entanglement for $N$ particles in this case is the {\it genuine multipartite entanglement }(GME), which is just $N$-particle entanglement. The main goal of quantum experiments is often proving that the quantum state created has a high level of multipartite entanglement or even showing that it has GME, which involves all the parties \cite{Toth2005Detecting,Sackett2000Experimental,Leibfried2005Creation,Kiesel2005Experimental,Walther2005ExperimentalOneWay,Haffner2005Scalable,Kiesel2007Experimental,Wieczorek2009Experimental,Prevedel2009Experimental,Song201710-Qubit,Monz201114-Qubit,Huang2011Multi-partite,Wang201818-Qubit,Friis2018Observation,Gong2019Genuine,Song2019Generation,Compact2021Pogorelov,Thomas2022Efficient,Cao2023Generation}. In the latter case, something qualitatively new has been created compared to experiments with fewer particles. 

In quantum metrology, it is known that  quantum entanglement is needed to surpass the classical limit in the precision of parameter estimation \cite{Pezze2009Entanglement}. Even weakly entangled bound entangled states can still be better than separable states \cite{Czekaj2015Quantum,Toth2018Quantum,Pal2021Bound}. It is also known that in order to reach a higher and higher metrological precision, higher and higher levels of multipartite entanglement is needed  \cite{Toth2012Multipartite,Hyllus2012Fisher}. This way, we can define metrologically useful $k$-particle entanglement, characterizing quantum states that are more useful metrologically than any quantum state with at most $(k-1)$-particle entanglement. States that do not possess metrologically useful $k$-particle entanglement form a convex set \cite{Toth2012Multipartite,Hyllus2012Fisher,Toth2020Activating}, similarly to other important sets of quantum states in entanglement theory, such as for instance separable states, which makes it possible to detect such entanglement with methods similar to the ones used in other areas of entanglement theory \cite{Horodecki2009Quantum,Guhne2009Entanglement,Friis2019}. Then, a quantum state of $N$-particles possesses metrologically useful GME, if it has metrologically useful $N$-particle entanglement. From these we can see that GME is needed to reach the maximal precision in parameter estimation. On the other hand, surprisingly, there exist pure states containing GME that are not more useful for metrology than separable states \cite{Hyllus2010Not}. Therefore, verifying the presence of metrologically useful entangled states is even more desirable than detecting entanglement without proving metrological usefulness \cite{Lucke2011Twin,Krischek2011Useful,Strobel2014Fisher}.

The question naturally arises: how could one activate those entangled states that are not useful for metrology, using a scheme that is relatively simple to implement in the lab, thus we will avoid applying a distillation step or local operations and classical communication (LOCC). We will consider multicopy metrology such that  only copies of the same party interact with each other, which is typical in activation schemes in quantum information science  \cite{Horodecki1999Bound,Activation2011Navascues,Palazuelos2012Superactivation,Shor2003Superactivation,Nawareg2017Superadditivity,Yamasaki2022Activation,Palazuelos2022Genuine,Morelli2022metrologyassisted}.  Looking for an optimal setup is challenging since the modeling of large quantum systems is needed.
 
We find that for $N$ qudits of dimension $d,$ there is a class of entangled states of at most rank-$d$ that become maximally useful compared to separable states in the limit of large number of copies. Surprisingly, the maximal usefulness is attained {\it exponentially fast} in the number of copies. Unexpectedly, the operators to be measured in order to verify the presence of metrologically useful GME turn out to be {\it simple correlations}. For $M $ copies, we need to measure a modest number of $M$ correlation terms. It is remarkable that this class contains metrologically useless, weakly entangled states that can even have an arbitrarily large overlap with product states. Thus, such states attain metrologically useful GME in the multicopy scenario, which contributes to the recent intensive research on activating entanglement using many copies \cite{Yamasaki2022Activation,Palazuelos2022Genuine,Morelli2022metrologyassisted}. A similar approach has been used to study the activation of genuine multipartite nonlocality \cite{Contreras-Tejada2021Genuine}. We will also consider how various relevant quantum states outside the subspace mentioned above perform in the multicopy scenario. Moreover, we will also show that embedding quantum states into higher dimensional spaces can also activate metrologically useful entanglement. We will support all our findings with powerful analytical and numerical methods that can describe multicopy metrology with large quantum systems. 

Our method might offer an approach for quantum metrology in the noisy, intermediate-scale quantum (NISQ) era \cite{Preskill2018Quantum}. In particular, we will suggest a procedure to force states into the desired class where metrologically useful GME activation is guaranteed, if they left the subspace due to an imperfect preparation or noise during the dynamics.

The paper is structured as follows.
In section~\ref{Sec:background}, we describe quantum metrology and the basic quantities to characterize the metrological performance of quantum states.
In section~\ref{Sec:setup}, we introduce the metrological scheme we use when considering multiple copies of quantum states.
In section~\ref{Sec:GME_activation}, we present the main result of the paper, that is, 
we identify a class of states for which metrologically useful GME activation is possible in the many copy limit. Moreover, we also provide the
measurements that need to be performed in order to reach the maximal metrological precision.
Then, we provide examples for quantum states inside and outside the above-mentioned subspace and evaluate their performance with our scheme. Furthermore, we also demonstrate the possibility of improving metrological performance simply by embedding quantum systems
into larger dimensions.
In section~\ref{sec:relation_to_bitflip}, we discuss a strategy that can be applied if the quantum states leave the subspace where metrologically useful GME activation is possible by making use of ideas from quantum error correction.
Finally, in section~\ref{sec:conclusion} we conclude the paper. 

\section{Quantum metrology}\label{Sec:background}

Before discussing our main results, we review some of the fundamental relations of quantum metrology \cite{Petz2008Quantum,Giovannetti2004Quantum-Enhanced,Paris2009QUANTUM,Demkowicz-Dobrzanski2014Quantum,Pezze2014Quantum,Toth2014Quantum,Pezze2018Quantum,Braun2018Quantum-enhanced,Sidhu2020Geometric,Polino2020Photonic,Barbieri2022Optical}. A basic metrological task in a {\it linear} interferometer is  estimating the small angle $\theta$ for a unitary dynamics 
\begin{equation}
\varrho_{\theta}=\exp(-i{\mathcal H}\theta)\varrho\exp(+i{\mathcal H}\theta),
\end{equation}
 where the Hamiltonian is the sum of {\it local} terms. In particular, for $N$-partite systems it is
\begin{equation}
{\mathcal H}={h}_1+{h}_2+...+h_N,\label{eq:bipartiteH}
\end{equation}
where $h_n$ for $n=1,2,\ldots,N$ are single-subsystem operators \footnote{For simplicity, we will denote by $h_n$ the operator acting on a single party as well as the operator acting on the entire system. The actual meaning can be inferred from the context.}. The precision is limited by the Cram\'er-Rao bound as \cite{Helstrom1976Quantum,Holevo1982Probabilistic,Braunstein1994Statistical,
Petz2008Quantum,Braunstein1996Generalized,Giovannetti2004Quantum-Enhanced,Paris2009QUANTUM,Demkowicz-Dobrzanski2014Quantum,Pezze2014Quantum,Toth2014Quantum,Pezze2018Quantum,Braun2018Quantum-enhanced,Sidhu2020Geometric,Polino2020Photonic,Barbieri2022Optical}
\begin{equation} \label{eq:cramerrao} 
\va{\theta}\ge\frac 1 {\nu \FQ[\varrho,{\mathcal H}]},
\end{equation}
where $\nu$ is the number of independent repetitions, and the quantum Fisher information, a central quantity in quantum metrology is defined by the formula \cite{Helstrom1976Quantum,Holevo1982Probabilistic,Braunstein1994Statistical,Petz2008Quantum,Braunstein1996Generalized}
\begin{equation}
\label{eq:FQ}
\FQ[\varrho,{\mathcal H}]=2\sum_{k,l}\frac{(\lambda_{k}-\lambda_{l})^{2}}{\lambda_{k}+\lambda_{l}}\vert \langle k \vert {\mathcal H} \vert l \rangle \vert^{2}.
\end{equation}
Here, $\lambda_k$ and $\ket{k}$ are the eigenvalues and eigenvectors, respectively, of the density matrix $\varrho,$ which is used as a probe state for estimating $\theta.$ From \EQ{eq:cramerrao} it can be seen that the larger the quantum Fisher information, the better precision we can achieve in parameter estimation. 
An efficient calculation method of the quantum Fisher information for large systems appears in \SUPP{app:efficient}.

We are interested in the ratio of the quantum Fisher information of a state and the maximum of the  quantum Fisher information for the same Hamiltonian for separable states, which we call the metrological gain for that particular unitary dynamics \cite{Toth2020Activating}
\be
g_{\mathcal H}(\varrho)=\frac{\FQ[\varrho,{\mathcal H}]}{\FQ^{({\rm sep})}(\mathcal H)}.\label{eq:gain}
\ee
The maximum for separable states is given as \cite{Ciampini2016Quantum,Toth2018Quantum,Toth2020Activating}
\begin{equation} 
\FQ^{({\rm sep})}(\mathcal H)=\sum_{n=1}^N [ \lambda_{\max}(h_n)-\lambda_{\min}(h_n) ]^2, \label{eq:seplim}
\end{equation}
where $\lambda_{\max}(X)$ and $\lambda_{\min}(X)$  denote the maximum and minimum eigenvalues of $X,$ respectively. Note that for qubits, if 
\be
h_n=\sum_{l=x,y,z} c_{l,n}\sigma_l,\label{eq:locham}
\ee
where $c_{l,n}$ are real numbers, and $|\vec c_n|=1,$ then $\FQ^{({\rm sep})}(\mathcal H)=4N$
\cite{Hyllus2010Not}. We also define the metrological gain optimized over all local Hamiltonians as
\be
g(\varrho)=\max_{\mathcal H}g_{\mathcal H}(\varrho).
\ee
If $g(\varrho)>1$ then the state is entangled and we call it also metrologically useful. 
If $h_n$ all have identical lowest and highest eigenvalues, then $g(\varrho)>k$ implies metrologically useful $(k+1)$-partite entanglement. If 
 $g(\varrho)>N-1$ then the state has metrologically useful GME, as discussed in \app{app:gain_multipartite_ent}.
In general, for quantum states $g(\varrho)\le N$ holds \cite{Toth2020Activating}. Note also that the metrological gain $g(\varrho)$ is convex in the quantum state \cite{Toth2020Activating}.

Finally, we mention that the variance and the Wigner-Yanase skew information defined as~\cite{Wigner1963INFORMATION}
\be
I_{\varrho}(\mathcal H)={\rm Tr}(\varrho \mathcal H^2)-{\rm Tr}(\sqrt{\varrho} \mathcal H\sqrt{\varrho} \mathcal H)\label{eq:WY}
\ee
provide upper and lower bounds, respectively, on the  quantum Fisher information as
\be
4\va{H}_{\varrho}\ge\FQ[\varrho,{\mathcal H}]\ge4I_{\varrho}(\mathcal H).\label{eq:FQI}
\ee
It is often easier to calculate $I_{\varrho}(\mathcal H)$ than $\FQ[\varrho,{\mathcal H}],$ which will be used in our derivations. 

\section{Multicopy scheme for activation}\label{Sec:setup}

In this section, we will consider metrology with several copies of the quantum state. First, we show that without interaction during the evolution, one cannot obtain an improvement in gain. Thus, it is not at all trivial that by adding new copies, the metrological gain will increase. Then, we study the setup based on an interaction between the copies of the same party. In this case, even the maximal gain can be achieved with very weakly entangled copies.

Let us consider $M$ copies of a quantum state, all undergoing a dynamics governed by the same Hamiltonian $\mathcal H.$ Then, for the quantum Fisher information we obtain 
\be
\FQ[\varrho^{\otimes M},\mathcal H^{\otimes M}]=M \FQ[\varrho,\mathcal H], \label{eq:EQM}
\ee
while the maximum for separable states also increases 
\be
\FQ^{({\rm sep})}(\mathcal H^{\otimes M})=M \FQ^{({\rm sep})}(\mathcal H).
\ee
Thus, the metrological gain does not change
\be
g_{\mathcal H^{\otimes M}}(\varrho^{\otimes M})=g_{\mathcal H}(\varrho).
\ee
Here, this statement is true for all schemes realizing an unbiased estimator in which there is no interaction during the quantum dynamics, however, after the dynamics the final states of the copies can be processed with any quantum circuit, even with collective measurements acting on several copies. Note that the gain remained the same since the quantum Fisher information given in \EQ{eq:EQM} increased $M$-fold, however, the performance of separable states also increased $M$-fold. 

\begin{figure}[h!]
\begin{center}\includegraphics[width=0.8\linewidth]{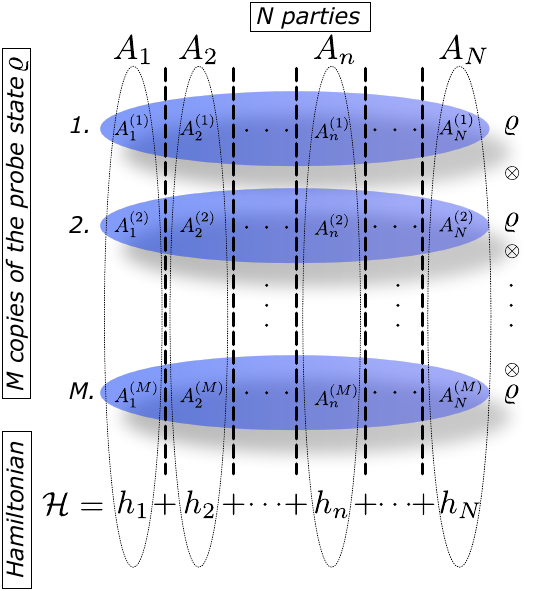}

\caption{Metrology with $M$ copies of an $N$-partite quantum state $\varrho$. Each horizontal ellipse represents a copy of the quantum state $\varrho$. The particles in the same vertical ellipse correspond to different copies of the same party $A_n$ and they can interact with each other during the evolution generated by the Hamiltonian $h_n$ for $n=1,2,...,N.$ However, there is no interaction between particles corresponding to different parties, which is stressed by vertical dashed lines separating the parties.} \label{fig:multicopy}\end{center}
\end{figure}

Hence, in order to increase the gain with the number of copies, we need to allow for interaction between the copies corresponding to the same party during the quantum dynamics. We consider $M$ copies of the $N$-partite state $\varrho$ acting on parties $A_n,$ as shown in \FIG{fig:multicopy}. 
The system consists of the subsystems $A_n^{(m)}$ for $m=1,2,...,M$ and $n=1,2,...,N.$ We calculate $g_{\mathcal H}[\varrho^{\otimes M},\mathcal H]$ with  local Hamiltonians of the form
\be
h_n=\otimes_{m=1}^{M}h_{A_n^{(m)}}\label{eq:h1}
\ee
for $n=1,2,...,N,$ as well as more general local Hamiltonians.  Extensive numerics show that $h_n$ of the form \eqref{eq:h1} can often reach the maximal metrological performance, which makes the implementation easier~\cite{Toth2020Activating}. In \SUPP{app:two_particle_terms}, we consider a different type of $h_n$ with only two-body correlations. 

We already know that any entangled bipartite pure state is maximally useful metrologically, in the limit of large number of copies \cite{Toth2020Activating}. However, what is the situation in the multiqubit case, relevant for quantum metrology with particle ensembles? What can we tell about the usefulness of mixed states?
\section{A metrologically useful GME activatable subspace}\label{Sec:GME_activation}
Here, we present the main result of the paper, that is, we identify a subspace in which all entangled multi-qudit quantum states can be made maximally useful. We also identify the measurements that have to be performed in order to achieve the maximal precision in metrology.

\begin{result}\label{obs:maxfqstates}
Entangled states of $N\ge 2$ qudits of dimension $d$ are maximally useful in the limit of large number of copies if they live in the 
\be
\{\ket{000..00},\ket{111..11},...,\ket{d-1,d-1,..,d-1}\}\label{eq:subspace}
\ee
subspace. The maximally achievable metrological usefulness is attained {\it exponentially fast in the number of copies}. States that can be transformed to this form with local unitaries have also the same property. 
\end{result}

{\it Proof.}---Let us consider 
\be
\varrho=\sum_{k,l=0}^{d-1} c_{kl} (\ket{k}\bra{l})^{\otimes N}, \quad  \mathcal H=\sum_{n=1}^N (D^{\otimes M})_{A_n}, \label{eq:definition_rho}
\ee
with $c_{kl}$ being the matrix elements of $\varrho$ in the basis from \EQ{eq:subspace} and $D={\rm diag}(+1,-1,+1,-1,...)$. To simplify the calculation, we use the mapping
\be\label{eq:mapping}
\varrho \rightarrow  \tilde \varrho=\sum_{k,l=0}^{d-1} c_{kl}\ket{k}\bra{l},\quad \mathcal H  \rightarrow  \tilde{\mathcal H}=N D^{\otimes M},
\ee
for which $\FQ[\varrho^{\otimes M},\mathcal H]=\FQ[\tilde \varrho^{\otimes M},\tilde{\mathcal H}]$ holds. We can bound the quantum Fisher information  as
\be
\FQ[\tilde \varrho^{\otimes M},\tilde{\mathcal H}]\ge4I_{\tilde \varrho^{\otimes M}}(\tilde{\mathcal H}),\label{eq:FQbound}
\ee
where the Wigner-Yanase skew information is
\be
I_{\tilde \varrho^{\otimes M}}(\tilde{\mathcal H})=N^2 [1-{\rm Tr}(\sqrt{\tilde \varrho} D \sqrt{\tilde \varrho} D )^M].
\label{eq:FQD}
\ee

In the limit of large number of copies,  if $[\sqrt{\tilde \varrho},D]\ne0$ then the skew information given in \EQ{eq:FQD} converges to the maximum. In this case, the state overcomes $4N,$ the separable limit of the quantum Fisher information given in \EQ{eq:seplim}, hence all such states are entangled. For $d\ge3,$ apart from the Hamiltonian $D,$ we should try other Hamiltonians that are obtained from $D$ by permuting its diagonal elements. If $\sqrt{\tilde \varrho}$ does not commute with one of these Hamiltonians, then, in the limit of large number of copies, the skew information with that Hamiltonian converges to the maximum, thus the state is entangled. If $\sqrt{\tilde \varrho}$ commutes with $D$ and with all the Hamiltonians obtained after permuting the diagonal elements, then $c_{kl}=0$ must hold for all $k\ne l.$ Such a state is a mixture of  product states.

The Wigner-Yanase skew information given in \EQ{eq:FQD} can be written out as follows for $d=2$
\begin{equation}\label{eq:final_res2}
 \frac{I}{N^2}=1-\Big[\frac{8 |c_{01}|^2 \sqrt{-c_{00}^2+c_{00}-|c_{01}|^2}+4 (c_{00}-1) c_{00}+1}{(1-2
   c_{00})^2+4 |c_{01}|^2}\Big]^M
\end{equation}
if $c_{01}\neq 0,$ otherwise $I=0.$ Moreover, if $c_{00}=c_{11}=1/2$ then equation~\eqref{eq:final_res2} can be simplified to 
\be
I(c_{01},N) = N^2[1 - (1-4|c_{01}|^2)^{M/2}].  \label{eq:c01N}
\ee 
$\qed$

\begin{figure}[h!]
\begin{center}\includegraphics[width=0.8\linewidth]{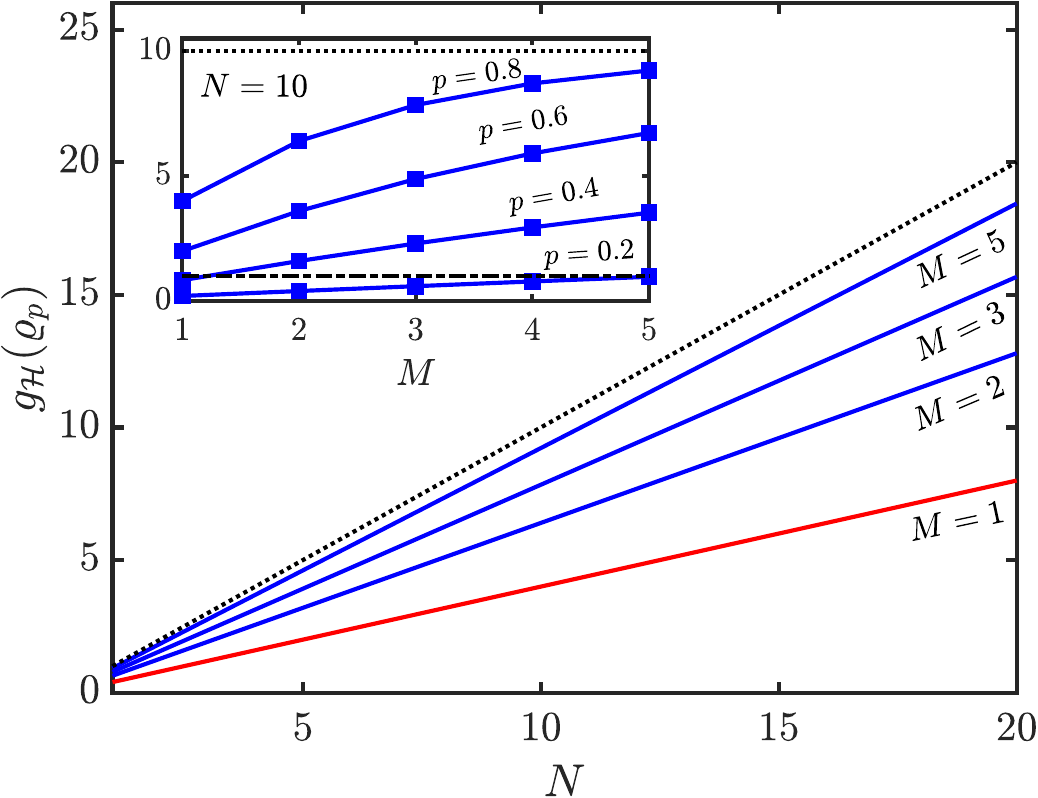}
\end{center}
\caption{Multicopy metrology with the noisy GHZ state given in \EQ{eq:rhop2} for $p=0.8.$ (solid) Lower bound on the metrological gain as a function of $N.$ $M$ denotes the number of copies. The state and the Hamiltonian are given in \EQ{eq:definition_rho}. We used the lower bound on the quantum Fisher information given in \EQS{eq:FQbound} and \eqref{eq:c01N}, where, for the case of $\varrho_p$ we have to set $c_{01}=p/2.$ Moreover, the separable limit is $\FQ^{({\rm sep})}=4N.$ (dotted) The maximal gain, $g_{\rm max}=N.$  (inset) The increase of the gain with the number of copies for $N=10$ for various values for $p.$ (dotted) The maximal gain $g_{\rm max}=10.$ (dashed dotted) Curve corresponding to $g=1.$} \label{fig:scaling}
\end{figure}

In Result~\ref{obs:maxfqstates}, we computed lower bounds on the quantum Fisher information. At this point, it could happen that the necessary measurements might be highly nonlocal operators. We will now show that, surprisingly, with simple operators it is possible to reach the maximal precision in parameter estimation. The operators to be measured are sums of {\it simple multiparty correlations}.

\begin{result}\label{obs:optmeas}
To achieve the maximal usefulness for the states appearing in \RES{obs:maxfqstates}, the following operator has to be measured \bea
\mathcal M&=&\sum_{m=1}^M Z^{\otimes (m-1)}\otimes Y \otimes Z^{\otimes (M-m)},\label{eq:MGHZnoisy}
\eea
where we define the multi-qubit operators acting on a single copy
\bea
Y&=&\begin{cases} \sigma_y^{\otimes N}&\text{ for odd $N$}, \nonumber\\\sigma_x\otimes\sigma_y^{\otimes (N-1)}&\text{ for even $N$}, \end{cases}\nonumber\\
Z&=&\sigma_z \otimes \mathds{1}^{\otimes (N-1)},\label{eq:MGHZnoisy2}
\eea
where $\sigma_x,\sigma_y,$ and $\sigma_z$ are the Pauli spin matrices.
By taking sufficiently many copies, the precision corresponding to the metrologically useful GME ($g=N$) can be approached fast and the required number of copies depends on how noisy the state is.
The proof is given in \app{app:optmeas}. 
\end{result}

Next, we will look at some consequences and applications of Result~\ref{obs:maxfqstates}, and we will also test the performance of our scheme for states living outside the subspace of Result~\ref{obs:maxfqstates}. Moreover, we will also provide some results concerning the single-copy case.

\subsection{Noisy GHZ states}

The method given in \RES{obs:maxfqstates} can be used to calculate the precision of the multicopy metrology with the state
\be
\varrho_p=p\ketbra{{\rm GHZ}}+(1-p)\frac{(\ketbra{0})^{\otimes N}+(\ketbra{1})^{\otimes N}}{2},\label{eq:rhop2}
\ee
where the Greenberger-Horne-Zeilinger (GHZ) state, playing a central role in quantum information science and quantum metrology, is \cite{Greenberger1990Bells}
\be
\ket{{\rm GHZ}}=\frac1{\sqrt 2}(\ket{0}^{\otimes N}+\ket{1}^{\otimes N}).
\ee
The state $\varrho_p$ is the one obtained after the particles of a GHZ state pass through a phase-noise channel, which is a relevant type of noise in many physical systems \cite{Nielsen2011Quantum}.

Let us see a concrete example with $M=2$ copies of the $3$-qubit case of the state $\varrho_p$ given in \EQ{eq:rhop2} with $p=0.8.$ Then, we have 
\be
\FQ[\varrho_p^{\otimes2},\mathcal H_{M=2}]=28.0976,
\ee
while in the single-copy case we have 
\be
\FQ[\varrho_p,\mathcal H_{M=1}]=23.0400.
\ee
Here, $\mathcal H_{M=2}$ and $\mathcal H_{M=1}$ are the Hamiltonians for the two-copy and single-copy cases, respectively, as defined in  \EQ{eq:definition_rho}. For these two cases, the upper bounds for the quantum Fisher information for separable states are
\be
\FQ^{({\rm sep})}(\mathcal H_{M=1})=\FQ^{({\rm sep})}(\mathcal H_{M=2})=12.
\ee
Hence, for the corresponding metrological gains
\be
g_{M=1}=1.92 < g_{M=2}=2.34
\ee
holds.

The state $\varrho_p$ turns out to be maximally useful in the limit of very many copies if $p>0$ as it is an entangled element of the set in Result~\ref{obs:maxfqstates}. For $\varrho_p,$ a lower bound on the quantum Fisher information can be obtained using the relation with the quantum Fisher information and the Wigner-Yanase skew information given in \EQ{eq:FQbound} and the inequality given in \EQ{eq:c01N}  with $c_{01}=p/2.$ We plot the lower bound on the metrological gain in \FIG{fig:scaling}. It can be seen that the lower bound approaches the theoretical maximum rapidly, as the number of copies is increasing.

\subsection{GHZ-like states with qudits of dimension $d>2$}
\RES{obs:maxfqstates} also implies that all entangled pure states of the form
 \be
\sum_{k=0}^{d-1}  \sigma_k \ket{k}^{\otimes N}\label{eq:Sch}
 \ee
 with $\sum_k \vert\sigma_k\vert^2=1$ are maximally useful in the limit of large number of copies. This can be seen as follows. Pure states given in \EQ{eq:Sch} form a subset of the states considered in \RES{obs:maxfqstates}. Among them, only states of the type $\ket{k}^{\otimes N}$are separable, therefore this is the only case where \RES{obs:maxfqstates} does not apply. 

Surprisingly, in the single-copy case not all entangled states of the type given in \EQ{eq:Sch} are useful. In particular, a single copy of the state given in \EQ{eq:Sch} for $d=2$ and $N\ge3$ is useful metrologically if and only if \cite{Hyllus2010Not}
\be
1/N<E:= 4\vert \sigma_0\sigma_1\vert^2.\label{eq:ab000111alpha}
\ee
For a short proof of this fact, which is more general than the one presented in \REF{Hyllus2010Not}, see \app{app:GHZlike}. In contrast, in the bipartite ($N=2$) case all entangled pure states are metrologically useful~\cite{Toth2020Activating} and to some extent the following result can be considered as a generalization of this fact.

\begin{result} 
\label{obs:allentpure}
All entangled pure states of the form given in \EQ{eq:Sch} with $\sum_k \vert\sigma_k\vert^2=1$ are useful for $d\ge3$ and $N\ge3.$ 
\end{result}
{\it Proof.}--- Let us see first the case of odd $d\ge 3$ and the block diagonal matrix
 \be\label{eq:hodd}
h^{({\rm odd})}_n={\rm diag}(1,X_{d-1}),
 \ee
 where $X_{d-1}$ is a $(d-1)\times (d-1)$ matrix with 1's in the antidiagonal and all other elements being 0. Hence,
 \be
 \FQ[\varrho,\mathcal H^{({\rm odd})}]=4N + 4N\vert\sigma_1\vert^2[N(1-\vert\sigma_1\vert^2)-1], \label{eq:fqodd}
 \ee
where the separable limit is $\FQ^{({\rm sep})}(\mathcal H^{({\rm odd})})=4N.$  With an appropriate permutation of the local basis states, from  $\mathcal H^{({\rm odd})}$ we can obtain a Hamiltonian for which $\sigma_k$ appears in the place of $\sigma_1$ in \EQ{eq:fqodd}. Hence, if for any $\sigma_k$
\be
0<\vert\sigma_k\vert^2 < (N-1)/N\label{eq.condsigma} 
\ee
holds then the given state is useful. If the state is entangled, then at least two of the $\sigma_k$ are nonzero, and one of them fulfills \EQ{eq.condsigma}.
 
Let us now consider the case of even $d\ge4$, with the block diagonal matrix
 \be
h^{({\rm even)}}_n={\rm diag}(1,1,X_{d-2}),
 \ee
 for which the quantum Fisher information is obtained as
  \be
\FQ[\varrho,\mathcal H^{({\rm even})}]=4N+4N(\vert\sigma_1\vert^2+\vert\sigma_2\vert^2)[N(1-\vert\sigma_1\vert^2-\vert\sigma_2\vert^2)-1]\label{eq:fqeven}
 \ee
holds. Similarly to the case of odd $d,$ with an appropriate permutation of the local basis states, from  $\mathcal H^{({\rm even})}$ we can obtain a Hamiltonian for which $\sigma_k$ and $\sigma_l$ with $l\ne k$ appear in the place of $\sigma_1$  and $\sigma_2,$ respectively, in \EQ{eq:fqeven}. Hence, if
 \be
0<\vert\sigma_k\vert^2 + \vert\sigma_l\vert^2 < (N-1)/N\label{eq.condsigma2} 
\ee
then the state is useful. If the state is entangled, then at least two 
of the $\sigma_k$ are nonzero. We have to examine the following cases.
 If the number of nonzero $\sigma_k$ is three or more, then two of the 
 $\sigma_k$ will fulfill \EQ{eq.condsigma2}. If only two of them are 
 nonzero then  we can consider a problem of odd $d$ with $d=3$ with 
 one $\sigma_k$ set to zero, and the state is useful. $\qed$
 
 \subsection{Activation of metrologically useful entanglement by embedding}
 
A surprising consequence of \RES{obs:allentpure} is that  all entangled states of the form given in \EQ{eq:Sch} are useful for $d=2,$ if we embed the qubits locally in qutrits, and consider a state as in  \EQ{eq:Sch} for $d=3$ by setting $\sigma_3=0.$ Thus, just by increasing the local dimension of the system, the states that have been found useless in \REF{Hyllus2010Not} can be activated and made useful.  It is also clear from the proof of \RES{obs:allentpure} that in this case, if we take $1/N=4\vert \sigma_0\sigma_1\vert^2$ [c.f.~\EQ{eq:ab000111alpha}] then for the asymptotic case of $N\rightarrow\infty,$ and for a single copy we obtain
\be
\FQ=5N, \quad\quad g=5/4,
\ee
by embedding, while $\FQ^{({\rm sep})}=4N.$ Here, in the spirit of the proof of \RES{obs:allentpure}, we 
consider $\mathcal H^{({\rm odd})}$ and also other Hamiltonians obtained from it after the local basis states are permuted. Thus, we improve metrological performance in the single-copy scenario, just by embedding the quantum states locally into a higher dimensional system. This can happen since after the embedding, new types of dynamics become possible that lead out from the original space. If such dynamics is allowed then the quantum state can outperform all separable states. Activation by embedding is related to activation by ancillas studied in \REF{Toth2020Activating}. The effect of ancillas have also been studied in \REF{Morris2020Entanglement}.

\subsection{Activation of metrologically useful GME from a non-useful state}
Let us take an entangled state of the form given in \EQ{eq:Sch} for $d=2$ and $N\ge3$ such that \EQ{eq:ab000111alpha} does not hold, which means that the state is non-useful $(g\le 1).$ We note that such a state can even be arbitrarily close to the fully polarized state. Despite being non-useful, according to \RES{obs:maxfqstates}, just by having several copies, we can reach metrologically useful GME with the above state ($g=N$). For further details see \SUPP{app:sclaing}. Thus, metrologically useful GME can be detected in these states \cite{Toth2012Multipartite,Hyllus2012Fisher}. 

The related problem of activating GME is at the center of attention in entanglement theory \cite{Yamasaki2022Activation,Palazuelos2022Genuine,Morelli2022metrologyassisted}. Note, however, that in our examples all states have been GME even in the single-copy case. It remains an important open problem whether metrologically useful GME can be activated using several copies of a quantum state without GME.

\begin{figure}[t]
  \begin{center}
  \includegraphics[width=0.8\linewidth]{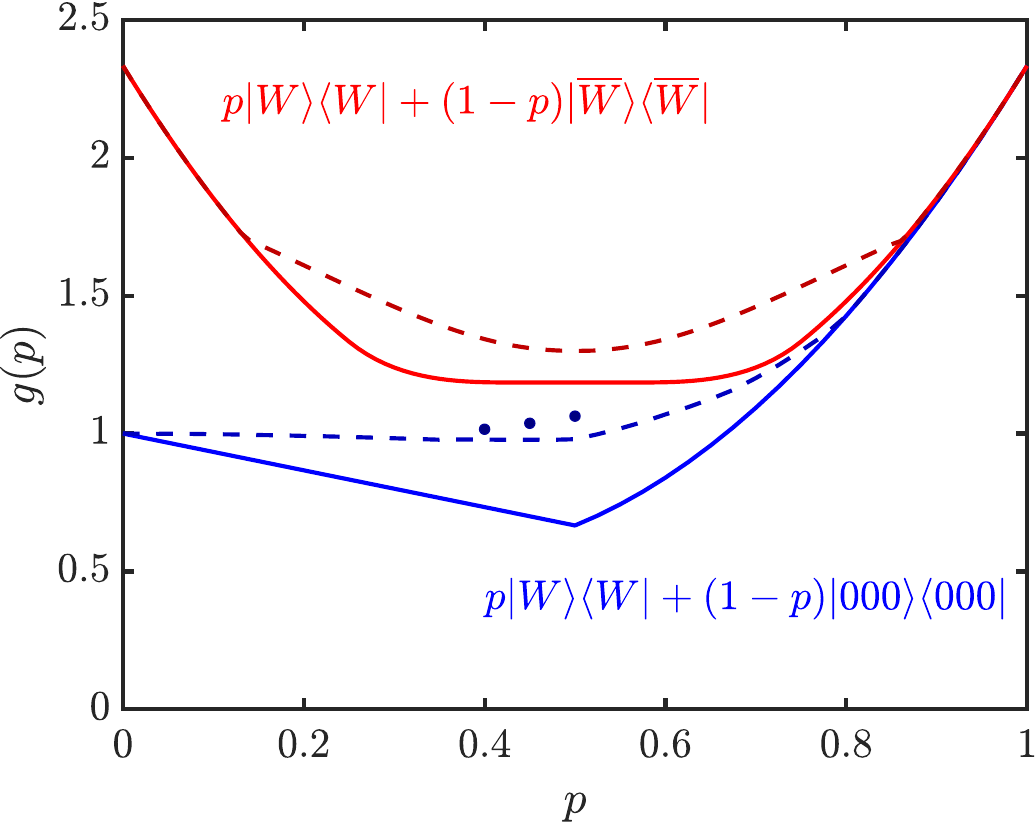}
  \end{center}
  \caption{Dependence of the metrological gain on $p$ for three-qubit mixtures of the W state and the $\overline{W}$ state and the three-qubit product state $\ket{000}$ given in  \EQ{eq:WWbarmixture}. 
  (solid) Single copy and (dashed) two copies. (top two curves) States with $r=0,$ that is, mixtures of the three-qubit $W$ and $\overline W$ states. (bottom two curves) States with $q=0,$ mixtures of the  three-qubit  $W$ state and $\ket{000}.$ (Individual dots) Three-copy case for $q=0$ and $p=0.4, 0.45,$ and $0.5.$ For all of them, $g>1$ holds.} \label{fig:W_Wbar}
  \end{figure}
\subsection{Mixtures of $W$ and  ${\overline W}$ states}

Now, in order to demonstrate that states living outside of the subspace described in \RES{obs:maxfqstates} can also be improved with our multicopy scheme let us consider a mixture of $W$ state defined as
\be
\ket{W}=\frac1{\sqrt{N}} (\ket{100...00}+\ket{010...00}+...+\ket{000...01}),
\ee
and the $\overline W$ state given as 
\be
\ket{\overline W}=\frac1{\sqrt{N}} (\ket{011...11}+\ket{101...11}+...+\ket{111...10}).
\ee
It is known that for three-qubit pure states, genuine multipartite entangled states can be equivalent to GHZ states or $W$ states under Stochastic LOCC (SLOCC) \cite{Dur2000Three}.  Thus, $W$ states represent a type of entanglement very different from that of GHZ states.

Let us examine first the $N=3$ case. Using the numerical methods of \REF{Toth2020Activating} that maximize the gain over Hamiltonians, we find that the optimal Hamiltonian for the $\ket W$ and the  $\ket{ \overline W}$  state is 
\be
\mathcal H=\sum_{n=1}^N \sigma_x^{(n)}.\label{eq:Hx}
\ee 
Now, let us now consider a broader family of states given by the mixture
\be
\varrho_{p,q}=p|W\rangle\langle W|+q|\overline W\rangle\langle\overline W|+r|000\rangle\langle000|,\label{eq:WWbarmixture}
\ee
where $p,q,r\ge 0$ and $p+q+r=1.$  First, let us examine the $r=0$ case. Such states have been studied for odd $N,$ since they are genuine multipartite entangled, still possess no multipartite correlations for $p=1/2$ \cite{Kaszlikowski2008Quantum,Bennett2011Postulates,Schwemmer2015Genuine}. We find that the metrological gain for states given in \EQ{eq:WWbarmixture} for $N=3$ is minimal for $p=1/2$ and the optimal Hamiltonian is
\be
\mathcal H=\sigma_z^{(1)}+\sigma_z^{(2)}-\sigma_z^{(3)},
\ee
which is translationally not invariant. 
For $p\approx 1$ and for $p\approx 0$ the maximal gain for two copies is the same as for a single copy, while for intermediate $p$ values the gain for two copies is larger, as can be seen in \FIG{fig:W_Wbar}. 

Next, let us consider states given in \EQ{eq:WWbarmixture} in the $q=0$ case. The maximal gain for two copies is always larger than for a single copy for $0<p<1$ as can be seen in \FIG{fig:W_Wbar}. We also tested the three-copy case for some $p$ values for which $g\le1.$ The metrological gain increases and states around $p=0.5$ are activated. Note that the state corresponding to $p=0.5$ is the state obtained from a four-qubit W-state, after a particle is lost. Thus, we can make such a state useful, if several copies are available.

For $N=4$ qubits and for a single copy, we find that the optimal Hamiltonian is of the type given in \EQ{eq:Hx}, for states $\ket W,$ $\ket{\overline W},$ and $\varrho_{1/2,1/2}.$ In such cases, the calculation of the quantum Fisher information can be simplified
as described in \SUPP{app:eff_single}.

Note that W states have been created experimentally, for instance, in trapped cold ions and photons \cite{Haffner2005Scalable,Eibl2004Experimental}, while quantum metrology has also been considered with the W and $\overline {\text{W}}$ states \cite{Li2023Quantum}. The type of noise considered can also be realized experimentally. We will discuss later the experiments creating GHZ states.

\subsection{Cluster states}

In this section, we study cluster states. They are highly entangled and can be used as a resource in measurement-based quantum computing \cite{Persistent2001Briegel,Raussendorf2001A}. Certain type of cluster states are known to be useless in the single-copy case in linear interferometers. Here, we show that surprisingly they remain useless even for multicopy metrology.

\begin{result} 
\label{obs:ring}
The ring cluster states for $N\ge5$ are not useful even in the limit of large number of copies.
\end{result}

\begin{figure}[t]
\begin{center}
\includegraphics[width=0.8\linewidth]{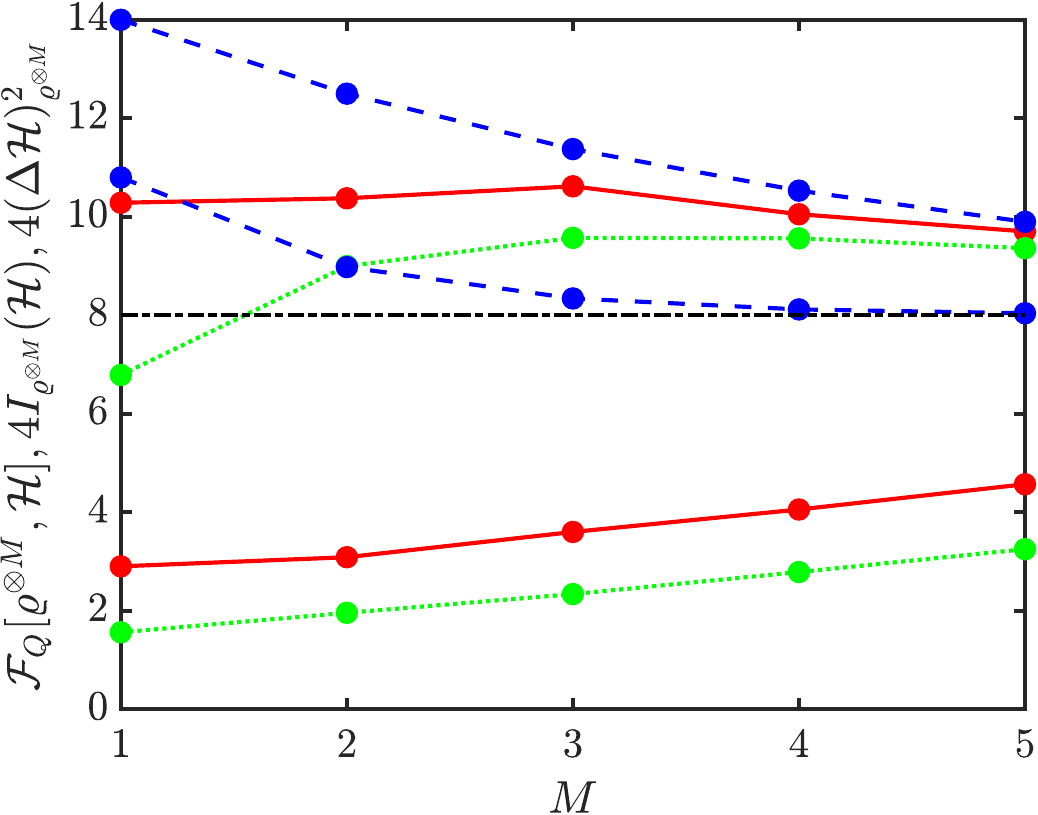}
\end{center}
\begin{center}\caption{The quantum Fisher information, and the
 upper and lower bounds given in \EQ{eq:FQI} as a function of the number of copies $M$ for the isotropic state of two qubits with $h_n=\sigma_z^{\otimes M}.$  (solid) $\FQ[\varrho_{\rm iso}^{\otimes M},\mathcal H],$ (dashed) the variance $4\va{\mathcal H}_{\varrho_{\rm iso}^{\otimes M}},$ (dotted) $4I_{\varrho_{\rm iso}^{\otimes M}}(\mathcal H)$ are plotted,  as well as (dashed dotted) the maximal quantum Fisher information  for separable states, $\FQ^{({\rm sep})}(\mathcal H).$ The noise parameter of the isotropic state is (top three curves) $p=0.75$ and (bottom three curves) $p=0.35.$} \label{fig:multicopy2}\end{center}
\end{figure} 

{\it Proof.}---Since the state is pure, the quantum Fisher information equals the variance times four. Let us consider an $N$-qubit ring cluster state $\ket{R_N},$ which is defined by the equations
\be
\sigma_z^{(n-1)}\sigma_x^{(n)}\sigma_z^{(n+1)}\ket{R_N}=\ket{R_N}
\ee
for $n=1,2,...,N,$ where $\sigma_z^{(0)}\equiv\sigma_z^{(N)}$ and $\sigma_z^{(N+1)}\equiv\sigma_z^{(1)}.$ For ring cluster states for $N\ge5$ all two-qubit reduced density matrices are the completely mixed state \cite{Hyllus2010Not}. Hence, for the reduced two-qubit states $\ex{\sigma_k \otimes \sigma_l}=0$ holds for $k,l=x,y,z,$ while for all reduced single-qubit states \ex{\sigma_l}=0 holds for $l=x,y,z.$ It is easy to see that this statement is also true for the multicopy case. Due to that, the variance equals the variance of the completely mixed state and can be obtained as
 \be
 \FQ[\varrho,\mathcal H]=4(\Delta \mathcal H)^2_{\varrho_{\rm cm}}=4\sum_{n=1}^N {\rm Tr}(h_n^2)/2^M,\label{eq:cmbound}
 \ee
 where we assumed ${\rm Tr}(h_n)=0,$ and $h_n$ are now $2^M\times2^M$ matrices representing operators acting on $A_n.$ The quantum Fisher information given in \EQ{eq:cmbound} is never larger than the separable limit in \EQ{eq:seplim} based on the well-known relation~\cite{popoviciu1935equations,bhatia2000better}
 \be
 \sum_{k=1}^{2^M} \lambda_k^2(h_n)/2^M \le  [ \lambda_{\max}(h_n)-\lambda_{\min}(h_n) ]^2/4,
 \ee
 where $\lambda_k(h_n)$ denote the eigenvalues of $h_n.$ $\qed$
 
\subsection{Two-qubit isotropic state}

Here, we consider $M$ copies of the two-qubit isotropic state, which is defined as  \cite{Horodecki1999Reduction,Horodecki1999General}
\be
\varrho_{\rm iso}(p)=p\ketbra{\Phi^+}+(1-p)\frac{\mathds{1}}4,
\ee
where $p$ is a noise parameter, and the maximally entangled state is 
\be
\ket{\Phi^+}=\frac 1 {\sqrt 2}(\ket{00}+\ket{11}).\label{eq:Phi+}
\ee
For the Hamiltonian, let us choose $h_n=\sigma_z^{\otimes M}$ for $n=1,2.$

The results are illustrated in \FIG{fig:multicopy2} for two different noise parameters. From this, we can see that there is an optimal number of copies above which the metrological performance does not improve. In the example with lower noise, the state is useful and the performance improves for $M=2,3$ copies, but the gain starts to decrease from $M=4$ copies. In the example with higher noise, the quantum Fisher information is increasing with $M$ but still does not overcome the separable limit.

To support the above observations from \FIG{fig:multicopy2}, we can obtain a general upper bound on the quantum Fisher information as
\be
\FQ\le 4 \ex{\mathcal H^2}=4\sum_n w_n^2+4 \sum_{n\ne n'} \ex{h_n \otimes h_{n'}}\le 4\left(\sum_n w_n\right)^2,\label{eq:fqineq}
\ee
where we used that  
\be
h_n^2=w_n^2\mathds{1}, \quad  \FQ^{({\rm sep})}(\mathcal H)=4\sum_n w_n^2
\ee
hold, where $w_n>0$ is some constant \footnote{It is sufficient to consider the case when $h_n^2\propto\mathds{1}$ for $n=1,2,..,N.$  If an operator $h_n$ is not of the above form, then we can consider the local Hamiltonians $h_n'$ with eigenvalues $\pm\lambda_{\max}(h_n')$ and eigenvectors identical to those of $h_n,$  which fulfill $(h_n')^2 \propto \mathds{1}.$ It has been shown that the metrological gain of at least one of the Hamiltonians $\mathcal H$ constructed from the local Hamiltonians above is larger or equal to that  of the original Hamiltonian \cite{Toth2020Activating}.}.  Then, we have 
\be
\ex{h_{n} \otimes h_{n'}}= (\ex{\sigma_z^{(n)}\otimes\sigma_z^{(n')}}_{\varrho})^M.
\ee
Thus, if $\ex{\sigma_z^{(n)}\otimes\sigma_z^{(n')}}_{\varrho}<1$ then the upper bound on the  quantum Fisher information given in \EQ{eq:fqineq} is going to the separable limit for large $M.$ 
 
 \section{Relation to the bitflip code}\label{sec:relation_to_bitflip}
Even if the initial state is ideally within the desired subspace of \RES{obs:maxfqstates}, in practice, due to imperfect preparation or noise during the dynamics it can eventually be outside, where metrologically useful GME activation is no longer guaranteed.
Here, we suggest a method to transform states not living in the subspace of \RES{obs:maxfqstates} into the subspace of \RES{obs:maxfqstates} based on ideas stemming from quantum error correction. 
This is to make sure that they can achieve maximal metrological performance with our scheme.

Error correction has been considered for quantum metrology \cite{Dur2014Improved,Kessler2014Quantum,Zhou2018Achieving}.
Using a bitflip error correcting code, a single qubit is stored in many qubits in the subspace mentioned in \RES{obs:maxfqstates} \cite{Shor1995Scheme,Steane1996Multiple-particle,Gottesman1996Class,Nielsen2011Quantum}. 
The syndrome measurements and restoring steps of the bitflip code can be used to move the state into the desired subspace even in our case. Note that we do not need to restore or protect a given quantum state, which is usually the case in error correction. We need only to obtain a quantum state that has a large metrological gain. The error correcting step mentioned above  can be carried out initially, and also throughout the dynamics. We can even avoid applying the correcting step. It is sufficient to employ a different Hamiltonian to states with different syndrome measurement results. 
We also analyze further relations to error correction in \SUPP{app:errcorr}.

GHZ states have been realized in trapped cold ions \cite{Sackett2000Experimental,Leibfried2005Creation,Monz201114-Qubit,Compact2021Pogorelov}, as well as error correction has also been carried out \cite{Chiaverini2004Realization,Schindler2011Experimental,Egan2021Fault-tolerant,Postler2022Demonstration}.  The GHZ states created have errors both in the diagonal and the off-diagonal elements of the density matrix when given in the computational basis. During the metrology, the main type of error is the decay of the off-diagonal elements \cite{Monz201114-Qubit}.  
Errors in the off-diagonal elements can be mitigated by multicopy metrology using $h_n=\sigma_z^{\otimes M},$ while errors in the diagonal elements can be overcome by the error correction scheme above.
Even if the error correcting steps are not applied, the multicopy scheme can increase the metrological gain. Superconducting circuits have also been used to create GHZ states \cite{Song201710-Qubit,Song2019Generation} and implement  error correction \cite{Reed2012Realization,Corcoles2015Demonstration,Ofek2016Extending,Zhao2022Realization,Krinner2022Realizing,Acharya2023Suppressing}. 
Error correction has recently been carried out in reconfigurable atom arrays \cite{Bluvstein2023Logical}. Our proposal might offer a viable information processing scenario in the NISQ era \cite{Preskill2018Quantum}, in which
simple error mitigation technics are needed \cite{Li2017Efficient,Temme2017Error,Endo2018Practical,Kandala2019Error,McArdle2019Error,Koczor2021Exponential,Yamamoto2022Error-Mitigated,Bharti2022Noisy,Rattew2021Efficient,Czarnik2021Error,Strikis2021Learning,Cerezo2021Applications}.

\section{Conclusion}\label{sec:conclusion}
In summary, we have presented an approach to activate metrologically useful multipartite entanglement. If the state is in a certain subspace, then, even if it was weakly entangled, it becomes maximally useful compared to separable states in the limit of  large number of copies, hence it will possess metrologically useful GME. Operations similar to the ones applied in error correction can be used to force the state into the desirable subspace. Our method involves simple measurements and can immediately be tested in present day quantum devices requiring moderate resources.
 
We have also shown that our scheme can improve the metrological performance of states living outside the above-mentioned subspace, like, for example for the two-qubit isotropic state. Moreover, we have also demonstrated the possibility of improving metrological performance by embedding quantum states locally into higher dimensions.

Deciding whether a quantum state is entangled or not is a hard problem, apart from small quantum systems. Deciding whether a quantum state possess metrologically useful entanglement for a given Hamiltonian is an easy task \cite{Horodecki2009Quantum}. However, deciding whether a state is metrologically useful in general needs an optimization over all local Hamiltonians \cite{Toth2020Activating}. It would be interesting to clarify whether this task is also computationally hard.

\section*{Acknowledgments}
We thank I.~Apellaniz, I.~L.~Egusquiza, G.~Giedke, C.~Klempt, J.~Ko\l ody\'nski, C.~D.~Marciniak, J.~J.~Meyer, G.~Muga, J.~Siewert, Sz.~Szalay, and G. Vitagliano for discussions.  We acknowledge the support of the  EU (COST Action CA15220, 
QuantERA MENTA, QuantERA QuSiED, QuantERA eDICT), the Spanish MCIU (Grant No.~PCI2022-132947), the Spanish Ministry of Science, Innovation and Universities and the European Regional Development Fund FEDER through, the Basque Government (Grant 
No.~IT1470-22), and the National Research, Development and Innovation Office NKFIH (Grants No.~K145927, No.~2019-2.1.7-ERA-NET-2020-00003, and No.~2019-2.1.7-ERA-NET-2021-00036).  We acknowledge the support of the Grant No.~PID2021-126273NB-I00 funded by MCIN/AEI/10.13039/501100011033 and by "ERDF A way of making Europe". We thank the "Frontline" Research Excellence Programme of the NKFIH (Grant No. KKP133827). We thank the project No. TKP2021-NVA-04, which has been implemented with the support provided by the Ministry of Innovation and Technology of Hungary from the National Research, Development and Innovation Fund, financed under the TKP2021-NVA funding scheme. We thank the Quantum Information National Laboratory of Hungary.  G.~T. thanks a  Bessel Research Award of the Humboldt Foundation. We thank the support from Foundation for Polish Science (IRAP project, ICTQT, contract no. 2018/MAB/5, co-financed by EU within Smart Growth Operational Programme). 

\appendix
\section{Metrological gain and multipartite entanglement}\label{app:gain_multipartite_ent}
  
We generalize the findings of \REF{Toth2012Multipartite,Hyllus2012Fisher} from qubits to higher dimensional systems. Let us assume that
$\lambda_{\max}(h_n)$ and $\lambda_{\min}(h_n)$ are identical for all parties $n=1,2,...,N.$ Then, based on \EQ{eq:seplim}, 
\begin{equation} 
\FQ^{({\rm sep})}(\mathcal H)=N \Lambda^2
\end{equation}
holds, where we define the difference between the largest and the smallest eigenvalue as 
\be
\Lambda=\lambda_{\max}(h_1)-\lambda_{\min}(h_1).
\ee 

Let us consider a pure state of the form
\be
\otimes_l \ket{\Psi_{k_l}},
\ee
where $\ket{\Psi_{k_l}}$ denotes a pure state of $k_l$ qudits and let us define the entanglement depth as $k=\max_l k_l.$
Then, for the quantum Fisher information we obtain that a series of inequalities
\bea\label{eq:FQLambda2}
\FQ[\varrho,{\mathcal H}] &\le& \Lambda^2 \sum_{l} k_l^2 \le \Lambda^2 [sk^2+(N-sk)^2] \le \Lambda^2 Nk,
\eea
where $s=\lfloor N/k \rfloor.$ The first inequality is based on \REF{Toth2012Multipartite,Hyllus2012Fisher}. For the second one, we need to prove that (page 68, \cite{Toth2021DScThesis})
\be
sk^2+(N-sk)^2\le Nk.\label{eq:sk2Nsk}
\ee
By substracting $sk^2$ from both sides of \EQ{eq:sk2Nsk}, we arrive at
\be
(N-sk)^2\le k(N-sk).\label{eq:sk2Nsk2}
\ee
\EQL{eq:sk2Nsk2} is evidently true, knowing that $0\le N-sk<k.$

Due to the convexity of the quantum Fisher information, the same bound holds also for quantum states that are mixtures of pure states with an entanglement depth at most $k.$
Hence, we obtain that 
\be
g > k
\ee
implies $(k+1)$-partite entanglement. From \EQ{eq:FQLambda2}, a more complicated, but somewhat stronger relation also follows
\be
g > [sk^2+(N-sk)^2]/N.
\ee

\section{Optimal measurements for the GME activatable subspace in Result~\ref{obs:maxfqstates}}\label{app:optmeas}
First, we will present the measurement operators explicitly for the $N$-qubit state 
\be
\varrho(p,q,r)=p\ketbra{{\rm GHZ}_q}+(1-p)[r(\ketbra{0})^{\otimes N}+(1-r)(\ketbra{1})^{\otimes N}],\label{eq:rhop2b}
\ee
where we choose a parametrization convenient for our derivation. Here, $0<p\le1$, $0\le r\le1,$ and the generalized GHZ state is \cite{Greenberger1990Bells}
\be
\ket{{\rm GHZ}_q}=\sqrt{q}\ket{000..00}+\sqrt{1-q}\ket{111..11},
\ee
where $0<q<1$ is real. 
Thus, the state $\varrho(p,q,r)$ turns out to be maximally useful in the limit of very many copies if $p>0$ and $0< q < 1,$ 
otherwise the state is a separable state. 

The error propagation formula, essentially, characterizes the uncertainty of the parameter estimation if we measure $\mathcal M$ as
\be
(\Delta \theta)^2_{\mathcal M}=\frac{\va{\mathcal M}}{\ex{i[\mathcal M,\mathcal H]}^2}.
\ee
The minimum is taken when $\mathcal M$ equals the symmetric logarithmic derivative, $\mathcal M_{\rm opt}$ which can be obtained from an explicit formula for a given $\varrho$ and $\mathcal H$ \cite{Demkowicz-Dobrzanski2014Quantum,Pezze2014Quantum,Toth2014Quantum,Pezze2018Quantum,Paris2009QUANTUM}. For $M=1,$ it is well known that \cite{Sackett2000Experimental,Leibfried2005Creation,Monz201114-Qubit}
\be
\mathcal M_{\rm opt} =\sigma_x^{\otimes N}.
\ee
On the other hand, for $M=2$ and $N=3$, for $q=r=1/2$ the optimal operator fulfills the following relations
\begin{align}
\langle \overline{0} \overline{0}| \mathcal M_{\rm opt} | \overline{0} \overline{0}\rangle  &=   0,
&\langle \overline{0} \overline{0} | \mathcal M_{\rm opt}| \overline{1} \overline{1}\rangle  &=   0,\nonumber\\
\langle \overline{0} \overline{1} |\mathcal M_{\rm opt} | \overline{1} \overline{1}\rangle  &=  -i,
&\langle \overline{0} \overline{0} |\mathcal M_{\rm opt} | \overline{0} \overline{1}\rangle  &= +i,
\end{align}
where $ \ket{\overline{0}}=\ket{000}$ and $\ket{\overline{1}}=\ket{111}.$ Based on these, one can obtain the optimal operator as 
\be
\mathcal M_{\rm opt}=\sigma_y \otimes \sigma_y \otimes \sigma_y \otimes \sigma_z \otimes \mathds{1} \otimes \mathds{1}
                  + \sigma_z \otimes \mathds{1} \otimes \mathds{1}  \otimes \sigma_y \otimes \sigma_y \otimes \sigma_y .\label{eq:Mopt}
\ee
Each correlation term is inverting the qubits at one of the copies and adds an additional phase conditioned on the state of the other copy.

Based on these observations, with straightforward algebra, it is easy to see that the operator, given by equation~\eqref{eq:MGHZnoisy} leads to the following error propagation formula
\be
(\Delta \theta)^2_{\mathcal M}=\frac{1/[4q(1-q)]+(M-1)p^2}{4MN^2p^2}.\label{eq:erporpqMN}
\ee
If the condition 
\be
1/[4q(1-q)]\ll(M-1)p^2
\ee
is fulfilled and $M\gg1$ holds, we have
\be
(\Delta \theta)^2_{\mathcal M}\approx \frac1{4N^2},\label{eq:Heisenberg}
\ee
which corresponds to the Heisenberg limit, which is the best achievable precision. 
Thus, measuring the operators, we can reach the Heisenberg limit for all the entangled states of the subspace defined in Result~\ref{obs:maxfqstates} in the multiqubit case.

Thus, for a given noise level $p$ and for a given value of the parameter $q,$ we need 
\be
M\approx \frac 1 {4q(1-q)p^2}
\ee
copies, which already leads to an almost optimal precision.  Much more copies will increase the precision somewhat, but will not lead to a much better performance. Moreover, the number of copies needed, $M,$ does not increase even if $N$ is increasing. Note that the operator given by equation \eqref{eq:MGHZnoisy}  is not the optimal one for $M\ge3.$ The optimal operator contains more correlation terms.

The $c$ matrix used in the proof of Result~\ref{obs:maxfqstates} can be given for the multiqubit state given in \EQ{eq:rhop2b} as
\be
c=\left(\begin{array}{cc}(1-p)r+pq & p\sqrt{(1-q)q} \\p\sqrt{(1-q)q} & (1-p)(1-r)+p(1-q) \end{array}\right).
\ee
This shows that we considered the most general $c$ with the only exception that $c_{01}$ is real and positive. The variable $c_{01}$ can always be made real and positive with an appropriate basis transformation
\be
\ket{1}\rightarrow \exp(-i\phi) \ket{1},
\ee 
where $\phi$ is an angle. Consequently,  the $\mathcal M$ operator corresponding to a state with a complex or a negative $c_{01}$ can be obtained by carrying out the inverse of such a basis transformation on $\sigma_x$ and $\sigma_y$ in the definition given in \EQ{eq:MGHZnoisy}.

Let us now consider the case of higher even dimensions $d.$ Let us consider a concrete example,
a state of higher dimensional qudits that is analogous to a GHZ state  given as  
\be
\ket{{\rm A}_d}=\frac1 {\sqrt d} \sum_{k=0}^{d-1} \ket{k}^{\otimes N},\label{eq:GHZd}
\ee
where $d$ is the dimensional of the qudits.
We can obtain $\mathcal M$ and all the $h_n$ operators  for metrology with the state given in \EQ{eq:GHZd} from the operators used in the qubit case using the substitution
\be
\sigma_l\rightarrow \oplus_{k=1}^{d/2} \sigma_l\label{eq:transform}
\ee
for $l=x,y,z.$ 
After the transformation, the operator to be measured is of the form given in \EQ{eq:MGHZnoisy}, where $X$, $Y$, and $Z$ are defined as 
\bea
Y&=&\begin{cases} (\oplus_{k=1}^{d/2} \sigma_y)^{\otimes N}&\text{ for odd $N$}, \nonumber\\
\oplus_{k=1}^{d/2} \sigma_x \otimes(\oplus_{k=1}^{d/2} \sigma_y)^{\otimes (N-1)}&\text{ for even $N$}, \end{cases}\nonumber\\
Z&=& (\oplus_{k=1}^{d/2} \sigma_z)\otimes\mathds{1}_d^{\otimes (N-1)},\label{eq:MGHZnoisy2_larged}
\eea
where $\mathds{1}_d$  is a  $d\times d$ identity matrix. 
Note that 
\be
\oplus_{k=1}^{d/2} \sigma_z=D,
\ee
where $D$ appears in the definition of the Hamiltonian given in \EQ{eq:definition_rho}. Thus, we really map the Hamiltonian given in Result~\ref{obs:maxfqstates} for qubit systems to the Hamiltonian given in Result~\ref{obs:maxfqstates} for qudit systems. 
The qubit states can also be mapped to qudit states as
\be
\ket 0 \rightarrow \ket{\tilde{0}}= \frac 1 {\sqrt{d/2} }(\ket 0 + \ket 2 +\ket 4 + ...),
\ee
and 
\be
\ket 1 \rightarrow \ket{\tilde{1}}=  \frac 1 {\sqrt{d/2} }(\ket 1 + \ket 3 +\ket 5 + ...).
\ee

Based on the transformations above, the multiqubit state given in \EQ{eq:rhop2} corresponds to the multiqudit state
\be
p\ketbra{A_d}+(1-p)\frac{(\ketbra{\tilde{0}})^{\otimes N}+(\ketbra{\tilde{1}})^{\otimes N}}{2}.\label{eq:GHZdnoisy2}
\ee
For the $\mathcal M$ and $\mathcal H$ obtained via \EQ{eq:transform} from the operators used for the qubit case, for the error propagation formula we obtain the same as in the qubit case we get for the state given in \EQ{eq:rhop2}, i.~e., \EQ{eq:erporpqMN} for $q=1/2.$ Thus, the setup reaches the Heisenberg limit given in \EQ{eq:Heisenberg} in the case of sufficiently large $M,$ if $p>0.$

Finally, let us consider the noisy state 
\be
p\ketbra{A_d}+(1-p)\frac{1}{d}\sum_{k=0}^{d-1}\ketbra{k}^{\otimes N}.\label{eq:GHZdnoisy}
\ee
The error propagation formula with the transformed $\mathcal M$ and $\mathcal H$ give the same result for the state given in \EQ{eq:GHZdnoisy} as for the state given in \EQ{eq:GHZdnoisy2}, since the operators appearing in the error propagation formula cannot distinguish the two states from each other. In particular, they cannot distinguish  the superposition of $\ket{n}$ and $\ket{n+2}$ from their mixture. Thus, even for the state given in \EQ{eq:GHZdnoisy}, the error propagation formula is given by \EQ{eq:erporpqMN} for $q=1/2$ and the setup reaches the Heisenberg limit given in \EQ{eq:Heisenberg} in the case of sufficiently large $M,$ if $p>0.$

\section{GHZ-like states}\label{app:GHZlike}

The condition for the usefulness of multipartite states  given in \EQ{eq:Sch} for $d=2$ and for $N\ge3$ has been presented already in \REF{Hyllus2010Not}, considering Hamiltonians given in \EQ{eq:locham}, where $c_{l,n}$ are real numbers, and $|\vec c_n|=1.$ For completeness, we present a very short proof, which also includes the case $|\vec c_n|\ne 1.$ 

\begin{result} 
A single copy of the state given in \EQ{eq:Sch} for $d=2$ is useful metrologically if and only if \EQ{eq:ab000111alpha} holds.
\end{result}

{\it Proof.}---Let us consider local Hamiltonians of the type given in \EQ{eq:locham},
where $\vert \vec c_n \vert=L_n.$ For this case, we obtain
\bea
\ex{\mathcal H^2}&=&\sum_n L_n^2+\left(\sum_n c_{z,n}\right)^2 -\left(\sum_n c_{z,n}^2\right),\nonumber\\
\ex{\mathcal H}^2&=&\left(\sum_n c_{z,n}\right)^2(1-E).
\eea
Let us first assume that $E>1/N.$ Then, we have the series of inequalities
\bea
g_{\mathcal H}&=&1+\frac{E\left(\sum_n c_{z,n} \right)^2-\sum_n c_{z,n}^2}{\sum_n L_n ^2} 
\le 1 +  \frac{(E-1/N)\left(\sum_n c_{z,n} \right)^2}{\sum_n L_n ^2}\nonumber\\
&\le&1 +  \frac{(E-1/N)\left(\sum_n L_n \right)^2}{\sum_n L_n ^2} 
\le N\cdot E.
\label{eq:FQGHZ}
\eea
In the first inequality, we used the inequality between the arithmetic and quadratic means for the $c_{z,n},$ $\left(\sum_n c_{z,n} \right)^2/N\le \sum_n c_{z,n}^2.$ In the third inequality, we used the same relation for $L_n.$  All inequalities are saturated if $c_{z,n}=L_n$ for all $n$ and they are all equal to each other. Thus, in this case we have 
$g_{\mathcal H}>1$ for a certain choice of local $\mathcal H,$ hence the state is useful.

Next, let us consider the $E\le 1/N$ case. Now, the first inequality in \EQ{eq:FQGHZ} is still valid and it leads to $g_{\mathcal H}\le1$  for any choice of local $\mathcal H,$ hence the state is not useful.

Thus, we obtain that the state is useful if and only if \EQ{eq:ab000111alpha} is fulfilled. $\qed$

Note also that for odd $N$, states given in \EQ{eq:Sch} for $d=2$ do not violate Bell inequalities with full correlation terms and two two-outcome observables per party if $E < 1/2^{N-1}$ \cite{Zukowski2002Do} [cf. \EQ{eq:ab000111alpha}].
\section*{References}
\bibliographystyle{iopart-num}
\bibliography{Bibliography2b}

\clearpage

\thispagestyle{empty}

\counterwithout{equation}{section}

\renewcommand{\thesection}{\Alph{section}}
\renewcommand{\thefigure}{S\arabic{figure}}
\renewcommand{\thetable}{S\arabic{table}}
\renewcommand{\theequation}{S\arabic{equation}}
\renewcommand{\theresult}{S\arabic{result}}

\stepcounter{mysection}
\stepcounter{myfigure}
\stepcounter{myequation}
\stepcounter{myresult}
\setcounter{page}{1}



\begin{center}
{\large \bf Supplemental Material for \\``Activation of metrologically useful genuine multipartite entanglement''}

\bigskip
R\'obert Tr\'enyi,\textsuperscript{1,2,3,4} \'Arp\'ad Luk\'acs,\textsuperscript{1,5,4} Pawe{\l}  Horodecki,\textsuperscript{6,7} \\ Ryszard Horodecki,\textsuperscript{6} Tam\'as V\'ertesi\textsuperscript{8} and G\'eza T\'oth\textsuperscript{1, 2, 3, 9, 4}
\smallskip

{\textsuperscript{1}\it  \small Department of Theoretical Physics,  University of the Basque Country UPV/EHU, \\P.O. Box 644, E-48080 Bilbao, Spain}

{\textsuperscript{2}\it \small EHU Quantum Center, University of the Basque Country UPV/EHU, \\Barrio Sarriena s/n, E-48940 Leioa, Biscay, Spain}

{\textsuperscript{3}\it  \small Donostia International Physics Center (DIPC), \\P.O. Box 1072, E-20080 San Sebasti\'an, Spain}

{\textsuperscript{4}\it  \small HUN-REN Wigner Research Centre for Physics, \\ P.O. Box 49, H-1525 Budapest, Hungary}

{\textsuperscript{5}\it  \small Department of Mathematical Sciences, Durham University,  \\Stockton Road, DH1 3LE Durham, United Kingdom}

{\textsuperscript{6}\it  \small International Centre for Theory of Quantum Technologies,\\ University of Gda\'nsk, Wita Stwosza 63, 80-308 Gda\'nsk, Poland}

{\textsuperscript{7}\it  \small Faculty of Applied Physics and Mathematics, National Quantum Information Centre,\\ Gda\'nsk University of Technology, Gabriela Narutowicza 11/12, 80-233 Gda\'nsk, Poland}

{\textsuperscript{8}\it  \small MTA Atomki Lend\"ulet Quantum Correlations Research Group, \\ HUN-REN Institute for Nuclear Research, \\ P.O. Box 51, H-4001 Debrecen, Hungary}

{\textsuperscript{9}\it  \small IKERBASQUE, Basque Foundation for Science,
E-48013 Bilbao, Spain}


\medskip
\medskip

\parbox[b][1cm][t]{0.85\textwidth}{\quad
The Supplemental Material contains some additional results. 
We show an efficient numerical method to calculate quantities for many copies of a quantum state.
We examine whether two-body interaction could be used instead of many-body interactions between the copies. 
We show relevant examples for the scaling of the precision with $N$ and $M.$
We present a simple algorithm that can be used to calculate the quantum Fisher information for the multicopy case for large systems, if the quantum Fisher information equals four times the variance. Surprisingly, there are relevant mixed states with this property.  We consider error mitigation schemes based on the ideas of the paper. Finally, we present a quantum state that is maximally useful and lives in the two-copy space. 
}
\bigskip
\bigskip
\bigskip
\bigskip
\bigskip
\end{center}

\bigskip
\bigskip
\bigskip
\bigskip
\bigskip

\setcounter{section}{0}
\renewcommand{\thesection}{Supplement~\Alph{section}}
\renewcommand{\theHsection}{appendixsection.\Alph{section}}
\renewcommand{\appendixname}{Supplement}
\section{Efficient numerics for many copies of a quantum state}\label{app:efficient}

We present efficient numerical methods to obtain the quantum Fisher information for large number of copies of a  bipartite state. The methods work even if a direct calculation would not be feasible. 

In order to obtain the quantum Fisher information given in \EQ{eq:FQ} for many copies, we have to compute the quantity
\be
 \langle k_1 |\langle k_2 |\langle k_3 | ... h_n ... \vert l_3 \rangle  |l_2 \rangle  |l_1 \rangle
\ee
for $n=1,2,...,N$  for large systems. $\ket{l_m}$ and $\ket{k_m}$ are all bipartite states, the eigenstates of the state we consider. This computation is possible if we note that if $\ket\Psi$ is a tensor product of two-particle states, then
\be
h_n \ket{\Psi}
\ee
is also a tensor product of two-particle states.

Let us now use the fact that the quantum Fisher information given in \EQ{eq:FQ} can be rewritten as 
\begin{equation}
\label{eq:FQ2}
\FQ[\varrho,{\mathcal H}]=4\ex{\mathcal H^2}-8\sum_{k,l}\frac{\lambda_{k}\lambda_{l}}{\lambda_{k}+\lambda_{l}}\vert \langle k \vert {\mathcal H} \vert l \rangle \vert^{2}.
\end{equation}
In the following, we assume that $\lambda_k>0$ only for $k=1,2,...,r,$ with $r$ being the rank of the state $\varrho.$

Based on \EQ{eq:FQ2}, we can write that 
\begin{equation}
\label{eq:FQN2}
\FQ[\varrho^{\otimes M},{\mathcal H}]=4\ex{\mathcal H^2}-8\sum_{\vec k,\vec l}\frac{\lambda_{\vec k}\lambda_{\vec l}}{\lambda_{\vec k}+\lambda_{\vec l}}
\vert \langle k_1 |\langle k_2 |\langle k_3 | ... {\mathcal H} ... \vert l_3 \rangle  |l_2 \rangle  |l_1 \rangle\vert^{2},
\end{equation}
where we use the notation
\be
\lambda_{\vec k}=\lambda_{k_1}\lambda_{k_2}\lambda_{k_3}...\lambda_{k_M}.
\ee
Here,  states with a zero eigenvalue do not contribute. It is sufficient to look at $\vec k$ and $\vec l$ that contain only $1,2,...,r.$  There are $r^{2M}$ terms in the sum in \EQ{eq:FQN2}.

We can simplify further our calculations if $h_{A^{(n)}}$ are all equal to each other 
\be 
h_{A_n^{(m)}}=h_{A_n},
\ee
for $n=1,2,...,N$ and $m=1,2,...,M.$ Due to permutational invariance, only the number of different eigenvalues matter, the order does not matter. Hence, we can write that 
\begin{equation}
\FQ[\varrho^{\otimes M},{\mathcal H}]=4\ex{\mathcal H^2}-8\sum_{\vec k\in \mathcal I(r),\vec l}{\mathcal P}_{\vec k}\times \frac{\lambda_{\vec k}\lambda_{\vec l}}{\lambda_{\vec k}+\lambda_{\vec l}}
\vert \langle k_1 |\langle k_2 |\langle k_3 | ... {\mathcal H} ... \vert l_3 \rangle  |l_2 \rangle  |l_1 \rangle\vert^{2},
\label{eq:FQN2n}
\end{equation}
where $\mathcal I(r)$ is the set of $M$-element vectors of $1,2,...,r$ with nondecreasing elements,  the coefficients based on the number of appearance of a certain term in the sum in \EQ{eq:FQN2} are given by the multinomial distribution
\begin{equation}
{\mathcal P}_{\vec k}=\binom{M}{m_1(\vec k ),m_2(\vec k ),...,m_r(\vec k )}.
\end{equation}
Here $m_l(\vec k )$ is number of $l$'s in $\vec{k}.$

Finally, we note that \EQ{eq:FQN2n} is a sum of positive terms that are also bounded from above. The average of positive bounded quantities can be approximated by an average obtained for a random sample. 

\section{Interaction between the copies via two-particle terms}\label{app:two_particle_terms}

Here, we consider $M$ copies of a state of $N$ qubits such that instead of $M$-particle correlations, two-particle correlations act between the copies. We show that for large number of copies, the metrological gain will be below $1$ and the state is not useful. This suggests that full correlations are needed.

Let us look at the correlations
\be
h_n=\sum_{m>m'} (\sigma_z)_{A_n^{(m)}} \otimes  (\sigma_z)_{A_n^{(m')}}\label{eq:Hsigmaz2}
\ee
for $n=1,2,...,N.$  Then we can make a mapping for the Hamiltonian 
\be
\mathcal H \rightarrow \tilde{\mathcal H}=N\sum_{m>m'}\sigma_z^{(m)} \otimes \sigma_z^{(m')}
\ee 
and for the state we use the mapping given in \EQ{eq:mapping} for $d=2.$ This way, we carry out our calculations with smaller matrices using
\be
\FQ[\varrho^{\otimes M},{\mathcal H}]=\FQ[\tilde \varrho^{\otimes M}, \tilde{\mathcal H}].
\ee
Then, one can prove the following series of inequalities
\bea
&&\FQ[\tilde \varrho^{\otimes M}, \tilde{\mathcal H}] \le 4N^2\va{M_z^2}_{\tilde \varrho^{\otimes M}} 
\le 4N^2 {\rm Tr}(\tilde \varrho^{\otimes M} M_z^4)\nonumber\\
&&\quad\le4N^2 {\rm Tr}(\ketbra{0}_x^{\otimes M}M_z^4)
=N^2(12M^2-8M).\label{eq:series}
\eea
In \EQ{eq:series}, the first inequality is based on that for any $A$ the inequality $\FQ[\varrho,A]\le 4\va{A}$ holds, and that we write the sum of two-body correlations as
\be
\sum_{m>m'} \sigma_z^{(m)}  \otimes  \sigma_z^{(m')}  =(M_z^2-M\mathds{1})/2,
\ee
where the $z$-component of the collective angular momentum is given as
\be
M_z=\sum_m \sigma_z^{(m)}.
\ee
In \EQ{eq:series}, the second inequality is based on that for any $A$ the inequality $\va{A}\le \ex{A^2}$ holds. In \EQ{eq:series}, the third inequality is based on that the product state $\tilde \varrho^{\otimes M}$  maximizing $M_z^4$ is $\ketbra{0}_x^{\otimes M}.$ In \EQ{eq:series}, the equality is based on simple analytical calculations. 

For the best metrological performance for separable states we have 
\begin{equation} 
\FQ^{({\rm sep})}(\tilde{\mathcal H})=N^2[ \lambda_{\max}(M_z^2/2)-\lambda_{\min}(M_z^2/2) ]^2. \label{eq:seplim2}
\end{equation}
The largest eigenvalue is
\be
\lambda_{\max}(M_z^2)=M^2.
\ee
The smallest eigenvalue is
\be
\lambda_{\min}(M_z^2)=\begin{cases}\text{$0,$} &\text{if $M$ is even,}\\ \text{$1,$} &\text{if $M$ is odd.}  \end{cases}
\ee
Based on these, for the metrological gain  
\be
g\le \begin{cases}\text{$(48M^2-32M)/M^4,$} &\text{if $M$ is even,} \\ \text{$(48M^2-32M)/(M^2-1)^2,$} &\text{if $M$ is odd,} \end{cases}
\ee
holds. The upper bound is decreasing with $M$ rapidly.  For $M\ge7$, the metrological gain will be below $1$ and the state is not useful metrologically.  

\section{Scaling properties}
\label{app:sclaing}

Here, we consider how the quantum Fisher information of the states appearing in \RES{obs:maxfqstates} scales with the number of parties. Moreover, we also analyze quantitatively the scaling properties of the state from equation~\eqref{eq:Sch} in detail.

Let us consider a family of quantum states given by the same coefficients $c_{kl}$ appearing in \EQ{eq:definition_rho} for all $N.$ For a constant $M,$ \EQS{eq:FQbound} and \eqref{eq:FQD} indicate that $\FQ\propto N^2$ holds. Hence, due to the Cram\'er-Rao bound in \EQ{eq:cramerrao}, the Heisenberg scaling, $\va{\theta}\propto1/N^2,$ can be reached. Moreover, the difference from the maximal quantum Fisher information is decreasing exponentially with $M.$ 

As a concrete example, let us consider the realistic scenario $c_{01}=\frac1 2 e^{-t/T},$ where $T$ is a time constant of the decay of $c_{01}.$ Then, from \EQ{eq:c01N} we obtain
\bea
I(c_{01},N) &\approx& N^2[1-(2t/T)^{M/2}]
\eea
for $t\ll T,$ using the first two terms for the Taylor expansion of the exponential. Thus, using several copies slows down considerably the decay of the metrological abilities with $t.$

Now, let us consider a family of states for which the coefficients $c_{kl}$ depend on $N$. In doing so, we will look at the behavior of the multicopy quantum states in the limit of large $N.$  Let us consider the state of the type given in \EQ{eq:Sch} for $d=2$ and $1/N=E:=4\vert \sigma_0\sigma_1\vert^2.$ Such a state has $g=1$ metrological gain, thus it is not useful metrologically~\cite{Hyllus2012Fisher}.  

From \RES{obs:maxfqstates} and \EQ{eq:final_res2}, we obtain for $M$ copies of the state 
\be
\FQ[\varrho^{\otimes M},\mathcal H]=4I_{\varrho^{\otimes M}}(\mathcal H)=4N^2[1-(1-1/N)^M],\label{eq:fqE0}
\ee
where the local Hamiltonians are 
\be
h_n=\sigma_z^{\otimes M}.
\ee
For this Hamiltonian, we have $\FQ^{({\rm sep})}(\mathcal H)=4N.$

Let us consider first a constant $M,$ not depending on $N.$ For that case and for $N\ll M,$ we obtain 
\be
(1-1/N)^M\approx 1.
\ee 
Hence, we have for the quantum Fisher information and the gain
\be
\FQ[\varrho^{\otimes M},\mathcal H]=4N^2,\quad\quad\quad g=N,
\ee
which corresponds to the Heisenberg limit. Then, for $N\gg M$ we obtain
\be
(1-1/N)^M\approx 1-M/N.
\ee 
Hence, for the $N\rightarrow\infty$ limit we have for the quantum Fisher information and the gain 
\be
\FQ[\varrho^{\otimes M},\mathcal H]=4NM,\quad\quad\quad g=M.
\ee

\begin{figure}[t]
\begin{center}
\includegraphics[width=0.43\textwidth]{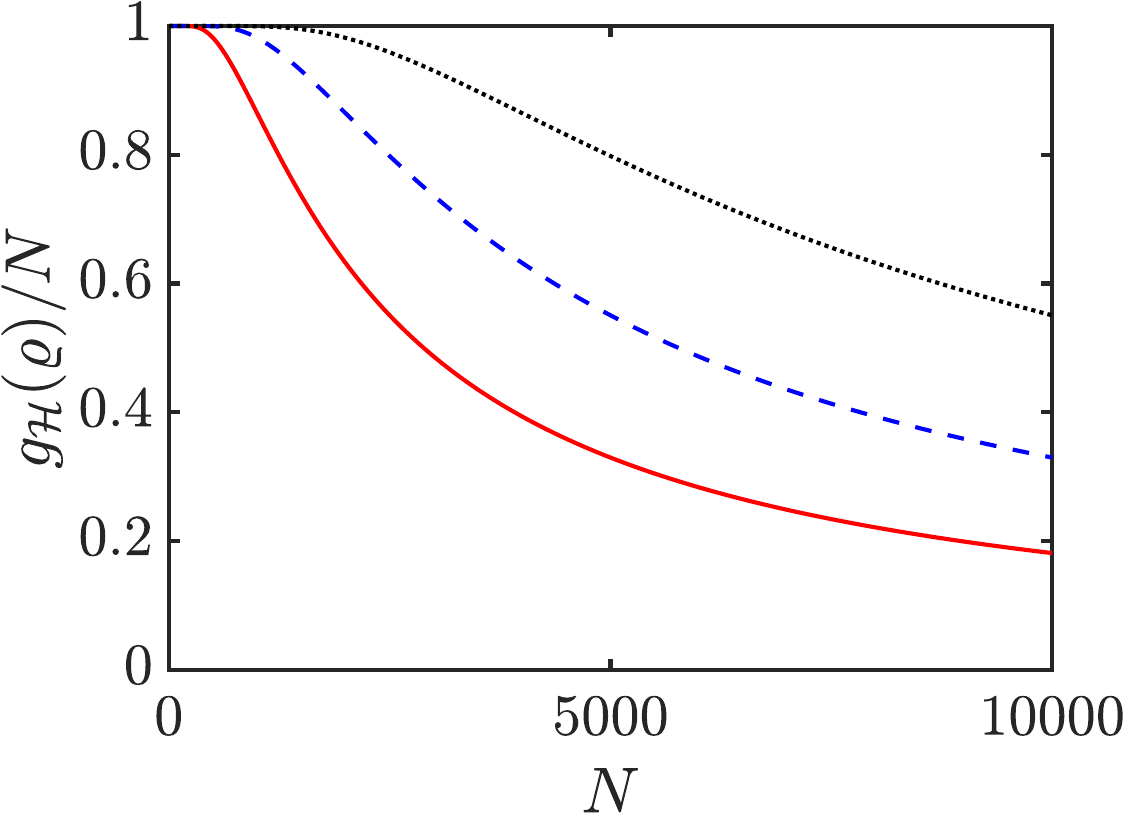} \hfill  \includegraphics[width=0.43\textwidth]{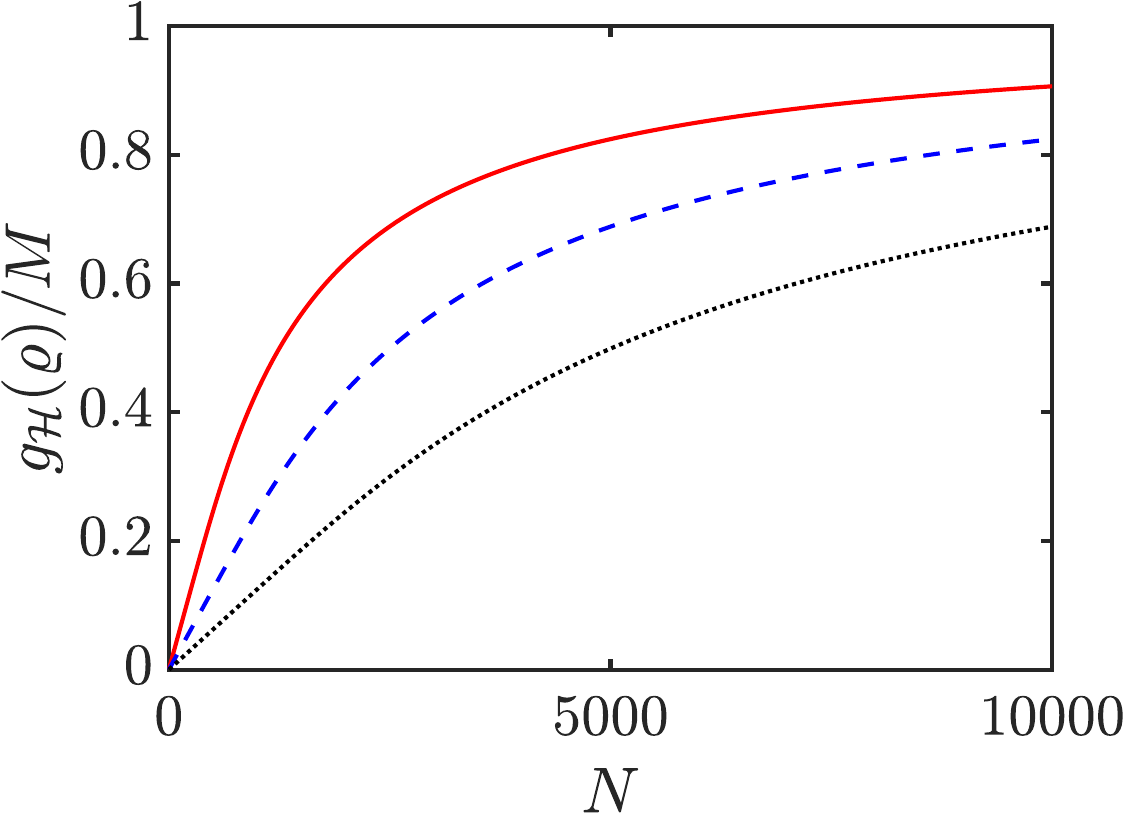}

\hskip0.5cm (a) \hspace{6.95cm} (b)
\end{center}
\caption{Dependence of the metrological gain on the particle number $N$ for (solid) $M=2000,$ (dashed) $4000$ and (dotted) $6000$ copies. (a) For $N\ll M$ we have $g=N.$ (b) For $N\gg M$ we have $g=M.$}\label{fig:Ndependence}

\end{figure}

In \FIG{fig:Ndependence}, it can be seen that for $N\ll M$ we reach the Heisenberg limit, while for $N\gg M$ we have at most the shot-noise scaling, {\it i.e.}, $\va{\theta}\propto 1/N.$

It is instructive to look at the case that $M$ depends on $N.$ If we set $M=N$ then in the $N\rightarrow\infty$ limit from \EQ{eq:fqE0} we obtain
\be
\FQ[\varrho^{\otimes M},\mathcal H]=4N^2 (1-1/e),\quad\quad g=N (1-1/e)\approx0.63N, \label{eq:FQN23}
\ee
where $e\approx2.7183$ is the basis of the natural logarithm, which corresponds to the Heisenberg scaling. We used the well known relation
\be
\lim_{N\rightarrow\infty}(1-1/N)^N=1/e.
\ee
If we set $N\ll M$ then in the $N\rightarrow\infty$ limit from \EQ{eq:fqE0}  we obtain
\be
\FQ[\varrho^{\otimes M},\mathcal H]=4N^2,\quad\quad\quad g=N,\label{eq:FQN22}
\ee
which is the Heisenberg limit. Finally, let us consider a case when the number of copies increases slower with the number of particles and set $M=\sqrt{N}.$ Then,  for the $N\rightarrow\infty$ limit we obtain
\be
\FQ[\varrho^{\otimes M},\mathcal H]=4N\sqrt{N},\quad\quad\quad g=\sqrt{N}.\label{eq:FQN2222}
\ee
Here we used the limit
\be
\lim_{N\rightarrow\infty}\frac{N^2[1-(1-1/N)^{\sqrt N}]}{N\sqrt{N}}=1.
\ee

Note that the probe state we considered was weakly entangled. For states that are more entangled, we can expect that the Heisenberg scaling can be reached with $M\ll N$ copies.

\section{Efficient calculations of the quantum Fisher information for large systems}\label{app:eff_single}

\begin{result} 
\label{obs:fqHkl}
If for the eigenvectors of the quantum state 
\be
\langle k \vert {\mathcal H} \vert l \rangle=0
\ee
holds for all $k,l$ for which $\lambda_k>0$ and $\lambda_l>0,$ then the quantum Fisher information can be calculated easily as
\begin{equation}
\FQ[\varrho,{\mathcal H}]=4\va{\mathcal H}.
\end{equation}
\end{result}

{\it Proof.}---Evidently, we have 
\begin{equation}
\langle \mathcal H \rangle=0.
\end{equation}
As we have seen in \EQ{eq:FQ2}, the quantum Fisher information given in \EQ{eq:FQ} can be rewritten as 
\begin{equation}
\label{eq:FQ22}
\FQ[\varrho,{\mathcal H}]=4\ex{\mathcal H^2}-8\sum_{k,l}\frac{\lambda_{k}\lambda_{l}}{\lambda_{k}+\lambda_{l}}\vert \langle k \vert {\mathcal H} \vert l \rangle \vert^{2},
\end{equation}
from which the statement follows. The advantage of \EQ{eq:FQ22} is that the summation is over cases when both $\lambda_k$ and $\lambda_l$ are nonzero. $\qed$

There are bound entangled state with this property \cite{Toth2018Quantum,Pal2021Bound}. This way, we can calculate the metrological performance of the state given in \EQ{eq:WWbarmixture} for $N\ge4$ if ${\mathcal H}=J_x,$ since for that state the condition of \RES{obs:fqHkl} is fulfilled. This is true even for the multicopy case, if the local Hamiltonians are 
\begin{equation}
h_n=\sigma_x^{\otimes M}.
\end{equation}

\section{Error mitigation scheme}
\label{app:errcorr}

Even without additional bitflip error correcting steps, it is worth to compare our scheme to error correction in quantum metrology via a concrete example \cite{Dur2014Improved,Kessler2014Quantum,Zhou2018Achieving}. In the usual error correction schemes, a general pure state is stored as a fully entangled multiqubit state. Let us consider the scheme of error correction assisted metrology from~\cite{Dur2014Improved} with a bitflip code such that a logical qubit corresponds to three physical qubits. Then we have the state
\begin{align}
\frac{1}{\sqrt2}(\ket{000\;000\;000}+\ket{111\;111\;111})
\end{align}
and the Hamiltonian
\begin{align}
H&=\sigma_{z}^{(1)}\sigma_{z}^{(2)}\sigma_{z}^{(3)}+\sigma_{z}^{(4)}\sigma_{z}^{(5)}\sigma_{z}^{(6)}+\sigma_{z}^{(7)}\sigma_{z}^{(8)}\sigma_{z}^{(9)}.\nonumber
\end{align}
For this case we also need error syndrome measurements to detect the error. Using the phase-flip code, we can suppress phase errors similarly, only a basis transformation is needed.

In contrast, in the case of multicopy metrology, we have $M$ copies of the state that are not entangled to each other initially and will become slightly entangled during the metrology. Let us consider the $M=3$ case with $h_n=\sigma_z^{\otimes 3}.$ Then, the state of the system for $N=3$ is
\begin{align}
\ket{\rm GHZ}^{\otimes 3}&=[\tfrac{1}{\sqrt2}(\ket{000}+\ket{111})]^{\otimes 3},
\end{align}
and the Hamiltonian is
\begin{align}
\mathcal{H}&=\sigma_{z}^{(1)}\sigma_{z}^{(4)}\sigma_{z}^{(7)}+\sigma_{z}^{(2)}\sigma_{z}^{(5)}\sigma_{z}^{(8)}+\sigma_{z}^{(3)}\sigma_{z}^{(6)}\sigma_{z}^{(9)}.\nonumber
\end{align}
Our approach improves the metrological performance without syndrome measurements. Moreover, it can suppress the effects of phase errors. In particular, the following result holds for $N\geq 2$ and $M=3$. 

\begin{result} Let $\varrho$ be a mixture of the following states that have phase error on at most one copy
\bea
\ket{\Psi_0}&=&\ket{{\rm GHZ}}\otimes\ket{{\rm GHZ}}\otimes\ket{{\rm GHZ}},\nonumber\\
\ket{\Psi_{1,\phi_1}}&=&\ket{{\rm GHZ}_{\phi_1}}\otimes\ket{{\rm GHZ}}\otimes\ket{{\rm GHZ}},\nonumber\\
\ket{\Psi_{2,\phi_2}}&=&\ket{{\rm GHZ}}\otimes\ket{{\rm GHZ}_{\phi_2}}\otimes\ket{{\rm GHZ}},\nonumber\\
\ket{\Psi_{3,\phi_3}}&=&\ket{{\rm GHZ}}\otimes\ket{{\rm GHZ}}\otimes\ket{{\rm GHZ}_{\phi_3}},\label{eq:phierror}
\eea
where
\be
\ket{{\rm GHZ}_{\phi}}=\frac{1}{\sqrt2}(\ket{0}^{\otimes N}+e^{-i\phi}\ket{1}^{\otimes N}),\nonumber
\ee
and $\phi_1$, $\phi_2$ and $\phi_3$ are arbitrary phase factors. We still obtain the maximal
\be
\FQ[\varrho,\mathcal H]=4N^2,\nonumber
\ee
where $\mathcal{H}$ is given by equation~\eqref{eq:bipartiteH}, such that $h_n=\sigma_z^{\otimes 3}$. For $N=3$, the optimal operator to be measured is 
\be
\sigma_z\otimes\mathds{1}\otimes \mathds{1} \otimes (\sigma_y)^{\otimes 6}+(\sigma_y)^{\otimes 6}\otimes\sigma_z\otimes\mathds{1}\otimes \mathds{1}+(\sigma_y)^{\otimes 3}\otimes\sigma_z\otimes\mathds{1}\otimes \mathds{1} \otimes(\sigma_y)^{\otimes 3}+\sigma_y^{\otimes 9}.\label{eq:optop}
\ee
\end{result}

{\it Proof.} Let us write $\varrho$ as
\be
\varrho=\sum_k p_k\ketbra{\Psi_k},
\ee
where $\ket{\Psi_k}$'s are chosen from the set in equation~\eqref{eq:phierror}. Then, for any $k,l$ we have
\be
\langle\Psi_k|\mathcal H|\Psi_l\rangle=0.
\ee
This can be seen as follows. Let us consider the tensor product form of $\ket{\Psi_k}$ from equation~\eqref{eq:phierror} as
\be
\ket{\Psi_k}=\otimes_{m=1}^3 \ket{\Psi_{k,m}},
\ee
where $\ket{\Psi_{k,m}}$'s are single copy states. We have
\be
\ex{h_n}=\prod_{m=1}^M \langle \Psi_{k,m}| \sigma_z | \Psi_{l,m}\rangle.
\ee
There is at most one erroneous copy in $\ket{\Psi_k}$ so for any $k, l$ there is always an $m$ such that $\ket{\Psi_{k,m}}=\ket{\Psi_{l,m}}=\ket{\rm{GHZ}}$. With this $m$, the relation $\langle \Psi_{k,m}| \sigma_z | \Psi_{l,m}\rangle=0$ holds. Hence,
\be
\langle\Psi_k|\mathcal H|\Psi_l\rangle=\bigg\langle\Psi_k\bigg|\sum_n h_n\bigg|\Psi_l\bigg\rangle=0.
\ee
Let us now consider the eigendecomposition of $\varrho$
\be
\varrho=\sum_k \lambda_k \ketbra{k}.
\ee
Then, any eigenvector for which $\lambda_k>0$ holds can be obtained as
\be
\ket{k}=\sum_l c_{kl} \ket{\Psi_l},
\ee
where $c_{kl}$ are some constants. Hence, for the eigenvectors in the range of $\varrho$ we have
\be
\langle k | \mathcal H | l \rangle=0.
\ee
Hence, it follows that \cite{Toth2018Quantum,Pal2021Bound}
\be
\FQ[\varrho,\mathcal H]=4\va{\mathcal H}_{\varrho}=4\ex{\mathcal H^2}_{\varrho}.\label{eq:FQVAReq}
\ee
Note that an optimization aiming at maximizing the quantum Fisher information within some set of quantum states often results in states satisfying \EQ{eq:FQVAReq} \cite{Toth2018Quantum,Pal2021Bound}. Based on \EQ{eq:FQVAReq}, due to the convexity of the quantum Fisher information and the concavity of the variance follows that for any $\ket{\Psi_k}$ having the properties mentioned above
\be
\FQ[\varrho,\mathcal H]=\sum_k p_k \FQ[\Psi_k,\mathcal H]
\ee
holds. If $\FQ[\Psi_k,\mathcal H]$ are maximal, so is $\FQ[\varrho,\mathcal H].$ Straightforward algebra shows that the operator given in \EQ{eq:optop} provides a precision that saturates the Cram\'er-Rao bound. $\qed$

Analogous arguments show that for any odd $M>3$ we can allow at most $(M-1)/2$ copies with phase errors and the quantum Fisher information will still be maximal. This is true even if we consider a mixture of the above states.

Note that a minimal experimental test is also possible with only two copies. In this case, we can consider a state that has no error and states that have an error on the first copy.

Note also that the application of the Hamiltionians $h_n=\sigma_{z}^{\otimes M}$ can be replaced by applying phase gates before and after the dynamics and using the Hamiltonian $h_n=\sigma_{z}\otimes\mathds{1}^{\otimes (M-1)}$ \cite{Dur2014Improved}. In this case, we can assume that the phase noise acts on the qubit that is coupled by $\sigma_z.$

\section{State living in the two-copy space}\label{app:two_copy_subspace}

In this section, we present a quantum state living in the two-copy space of a bipartite system. We denote the two copies $AB$ and $A'B'.$

Let us consider the quantum state
\be
\varrho_p=p\ketbra{\Psi^+}^{\otimes2}+(1-p)\ketbra{\Phi^+}^{\otimes2},\label{eq:rhoppp}
\ee
where $\ket{\Phi^+}$ is defined as in \EQ{eq:Phi+}, and 
\be
\ket{\Psi^+}=\frac{1}{\sqrt 2}(\ket{01}+\ket{10}).
\ee
For state $\varrho_p$ and the Hamiltonian
\be
H=\sigma_z^{A}\sigma_z^{A'}+\sigma_z^{B}\sigma_z^{B'}
\ee
for any $p$ the metrological gain is obtained as
\be
g=2.
\ee
The state $\varrho_p$ given in \EQ{eq:rhoppp} is equivalent to the state that is the mixture of the maximally entangled state
\be
\frac{1}{2}\sum_{n=0}^{3} \ket{n}_{AA'}\ket{n}_{BB'},
\ee
and state obtained from it after a basis transformation
\be
\frac{1}{2}\sum_{n=0}^{3} \ket{n}_{AA'}\ket{3-n}_{BB'}.
\ee
It is instructive to compare the state given in \EQ{eq:rhoppp} to 
\be
\varrho_p=p\ketbra{\Psi^+}+(1-p)\ketbra{\Phi^+},\label{eq:rhoppp222}
\ee
which is nonentangled and metrologically not useful for $p=1/2.$

\clearpage
\eject


\end{document}